# Topological resonance behaviors of surface acoustic waves under a surface liquid-layer loading and sensing applications

Bowei Wu [1], Tingfeng Ma [1,*], Shuanghuizhi Li [1], Xiang Fang [1], BoyueSu [1], Peng Li [2], Zhenghua Qian [2] , Rongxing Wu [3], Iren Kuznetsova [4], Vladimir Kolesov [4]

[1] Zhejiang-Italy Joint Lab for Smart Materials and Advanced Structures, School of Mechanical Engineering and Mechanics, Ningbo University, Ningbo 315211, China;

[2] The State Key Laboratory of Mechanics and Control of Mechanical Structures, College of Aerospace Engineering, Nanjing University of Aeronautics and Astronautics, Nanjing 210016, China;

[3] Department of Architectural Engineering, Ningbo Polytechnic, Ningbo 315800, China;

[4] Kotelnikov Institute of Radio Engineering and Electronics of RAS, Moscow 125009, Russia.

## Abstract

In this work, topological resonance behaviors of surface acoustic waves (SAW) under a surface liquid-layer loading are investigated. By revealing influences of the liquid-layer loading on wave velocity of SAW and topological indices (Berry curvature and Chern number) of topological interface-modes, a topological resonance peak with a high Q-factor is obtained based on couplings of a topological interface-mode waveguide and a resonant cavity under a surface liquid-layer loading. The results show that the degree of spatial-inversion-symmetry breaking resulting from structure parameters has an obvious influences on the topological resonance Q-factor, while the influences of the thickness of the liquid-layer loading on that is weak. It is worth noting that the topological resonance frequency is significantly sensitive to the liquid parameters. Based on that, a novel topological-resonance SAW liquid-phase sensor is proposed. Furthermore, sensing performances of this kind of sensor are simulated, which are used to sensing the concentration of hemoglobin, albumin, NaCl and NaI in aqueous solutions, and high sensitivities and Q-factors are obtained. The results presented in this paper can provide an important basis for the realization of highly sensitive and stable SAW micro-liquid-sample biomedical sensors in the future.

* Corresponding authors.

Email address: matingfeng@nbu.edu.cn.

# 1. Introduction

In recent years, surface acoustic wave (SAW) devices are playing increasingly important roles in the fields of communications and sensings [1–5]. It is notable that SAW technologies have been applied in liquid-phase biomedical detections [6–9] and microfluidics [10–12]. However, for traditional SAW sensors, the energy concentration degree of SAW during the transmission process is relatively low, which limits its sensitivity. Besides, it is challenging to accurately and conveniently control changes of the transmission direction of SAW. These limitations hinder the development of high-performance SAW sensors obviously.

For the above reasons, some sensors based on phononic crystals (PnC) have been designed to detect concentrations of solutions [13–15], which primarily utilize defect-state transmission of acoustic or elastic waves in phononic crystals to control conveniently the energy concentration and transmission direction [16–21]. However, for PnC sensors based on defect states, the range of the operational frequency is narrow, which limits the versatility of detected objects. Besides, the energy dissipation of wave transmission in PnC leads to a low resonance quality factor (Q-factor), which weakens the stability of the sensor obviously [22]. Meanwhile, the lack of sufficient manufacturing precision can also increase energy dissipation, further affecting the detection stability of the sensor.

For the last few years, topological transmissions with less energy dissipation have attracted widespread attentions in the fields of electromagnetic wave [23–25], acoustic waves [26–29], and elastic waves [30–35], due to abilities of suppressing backscattering and being immune to defects. The energy concentration of SAW topological transmissions is superior obviously to that of traditional SAW transmissions [36–41]. Based on the coupling between the topological waveguide and the resonant cavity, topological resonance with high Q-factor has been realized [42, 43], which is suitable for sensing applications.

In the field of biomedical detection, most sample is liquid. However, the liquid-layer sample brings a complex loading conditions to topological-resonance SAW devices. Currently, influences of the surface liquid-layer loading on the energy-band structure and topological characteristics of SAW topological devices are unclear, which hinders the development of high-sensitivity micro-liquid-sample sensors based on SAW topological-resonances. Therefore, exploring and realizing the topological resonance behavior of SAW covered by a liquid-layer loading is crucial for the development of high-performance micro-liquid-sample biomedical sensors.

In this work, influences of the liquid-layer loading on wave velocity of SAW and topological indices (Berry curvature and Chern number) of topological interface-modes are revealed, and then a topological resonant peak with a high Q-factor is obtained based on couplings of a topological interface-mode waveguide and a resonant cavity under a surface liquid-layer loading, on this basis, high-sensitivity and high Q-factor have been achieved for micro-liquid-sample sensors.

The structure of this paper is as follows: In Section 2 (Model), the effects of liquid-layer loadings on the transmission velocity of SAW and the realization of topological phase transitions are presented; In Section 3, topological resonance behaviors of SAW under a liquid-layer loading based on topological waveguide-resonator couplings are investigated and realized; In Section 4, Sensing applications based on the topological resonance behaviors of SAW are presented; The conclusion is given in Section 5.

## 2. Model

A theoretical model is constructed to examine the influences of parameters of liquid-layer loadings on the wave velocity of Rayleigh SAW. The model is shown in Fig.1. The piezoelectric substrate of a half-space with $x_3>0$ is considered and the fluid with a thickness of *h* occupies a space where $x_3<0$.

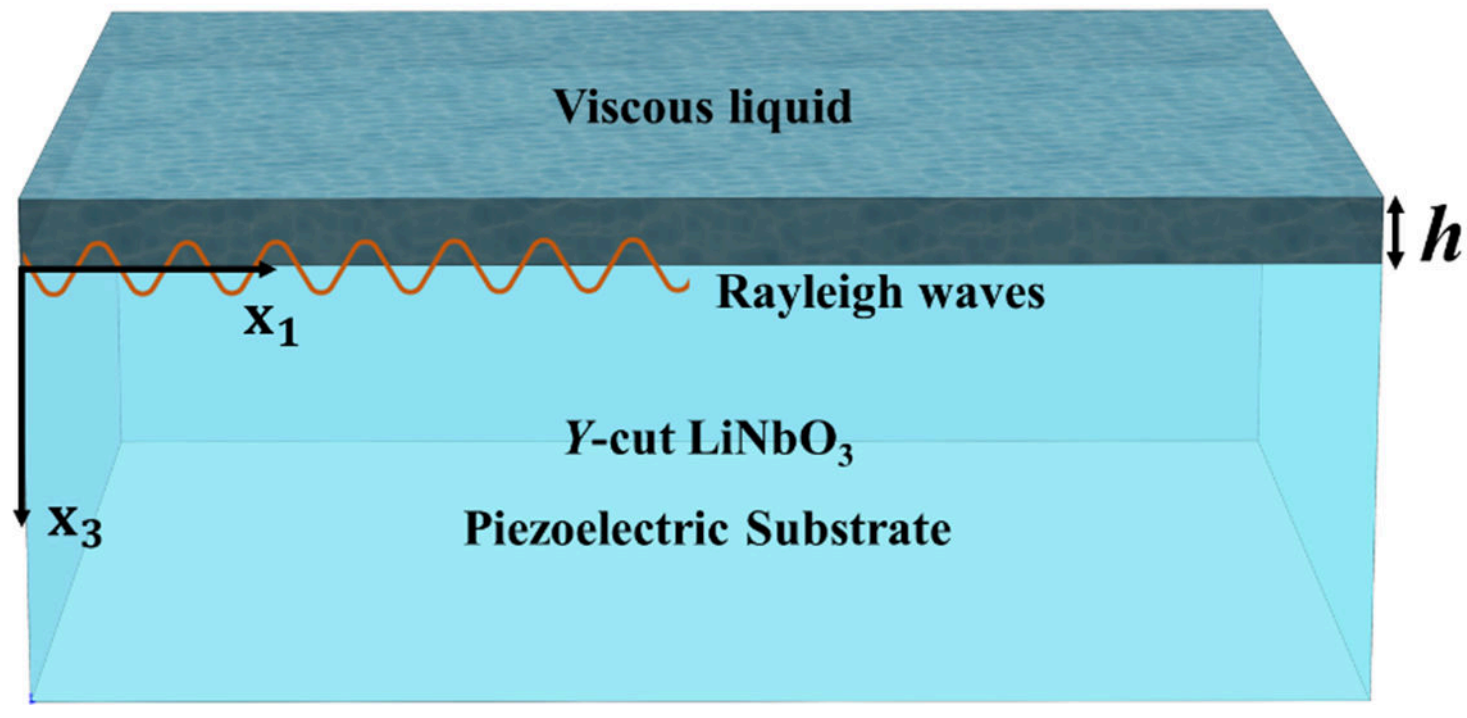

**Fig. 1.** Schematic diagram of propagations of Rayleigh SAW in a $LiNbO_3$ substrate under a liquid-

layer loading

Assuming Rayleigh waves travel along *x* direction, the displacement components and electric potential are respectively [44–46]:

$$u_1 = u_1(x_1, x_3, t),\ u_2 = 0,\ u_3 = u_3(x_1, x_3, t),\ \phi = \phi(x_1, x_3, t), \tag{1}$$

where $x_i\ (i=1,\ 3)$ denotes the *x* and *z* directions, respectively; $\phi$ is the electric potential. $u_i\ (i=1,\ 2,\ 3)$ are displacements in *x*, *y* and *z* directions, respectively.

The equilibrium equations of elasticity without body forces are

$$\sigma_{ji,j} = \rho\ddot{u},\ D_{i,i} = 0,\ (i, j = 1,2,3), \tag{2}$$

where $\sigma_{ji,j}$, $D_{i,i}$ and $\rho$ are the stress, electric displacement and mass density, respectively.

For an anisotropic and elastic solid, the coupled constitutive relations are

$$\begin{aligned} \sigma_i &= c_{ik} S_k - e_{ki} E_k, \\ D_i &= e_{ik} S_k + \varepsilon_{ik} E_k, \end{aligned} \tag{3}$$

where, $S_k$, $E_k$ are the strain and electric field components, respectively; $c_{ik}$, $e_{ik}$ and $\varepsilon_{ik}$ are the elastic, piezoelectric, and dielectric coefficients, respectively. The strain *S* and the electric field *E* are respectively

$$\begin{aligned} \mathrm{S}_{ik} &= (u_{i,k} + u_{k,i}) / 2, \\ E_i &= -\phi_{,i}. \end{aligned} \tag{4}$$

Substitution of Eq. (4) into Eq. (3) yields

$$\begin{bmatrix} \sigma_{11} \\ \sigma_{22} \\ \sigma_{33} \\ \sigma_{23} \\ \sigma_{31} \\ \sigma_{12} \end{bmatrix} = \begin{bmatrix} c_{11} & c_{12} & c_{13} & c_{14} & 0 & 0 \\ c_{21} & c_{21} & c_{23} & c_{24} & 0 & 0 \\ c_{31} & c_{31} & c_{33} & c_{34} & 0 & 0 \\ c_{41} & c_{42} & c_{43} & c_{44} & 0 & 0 \\ 0 & 0 & 0 & 0 & c_{55} & c_{56} \\ 0 & 0 & 0 & 0 & c_{65} & c_{66} \end{bmatrix} \begin{bmatrix} u_{,1} \\ v_{,2} \\ w_{,3} \\ v_{,3} + w_{,2} \\ u_{,3} + w_{,1} \\ v_{,1} + u_{,2} \end{bmatrix} - \begin{bmatrix} 0 & e_{21} & e_{31} \\ 0 & e_{22} & e_{32} \\ 0 & e_{23} & e_{33} \\ 0 & e_{24} & e_{34} \\ e_{15} & 0 & 0 \\ e_{16} & 0 & 0 \end{bmatrix} \begin{bmatrix} -\phi_{,1} \\ -\phi_{,2} \\ -\phi_{,3} \end{bmatrix},$$

$$\begin{bmatrix} D_1 \\ D_2 \\ D_3 \end{bmatrix} = \begin{bmatrix} 0 & e_{21} & e_{31} \\ 0 & e_{22} & e_{32} \\ 0 & e_{23} & e_{33} \\ 0 & e_{24} & e_{34} \\ e_{15} & 0 & 0 \\ e_{16} & 0 & 0 \end{bmatrix}^{\mathrm{T}} \begin{bmatrix} u_{,1} \\ v_{,2} \\ w_{,3} \\ v_{,3} + w_{,2} \\ u_{,3} + w_{,1} \\ v_{,1} + u_{,2} \end{bmatrix} + \begin{bmatrix} \varepsilon_{11} & 0 & 0 \\ 0 & \varepsilon_{22} & \varepsilon_{23} \\ 0 & \varepsilon_{32} & \varepsilon_{33} \end{bmatrix} \begin{bmatrix} -\phi_{,1} \\ -\phi_{,2} \\ -\phi_{,3} \end{bmatrix}. \tag{5}$$

Here, wave propagations in *y* direction are not considered, so the equilibrium differential equation can be written as

$$
\begin{aligned}
&(c_{11}\frac{\partial^2 u_1}{\partial {x_1}^2}+c_{55}\frac{\partial^2 u_1}{\partial {x_3}^2})+(c_{13}+c_{55})\frac{\partial^2 u_3}{\partial x_3 \partial x_1}+(e_{15}+e_{31})\frac{\partial^2 \phi}{\partial x_3 \partial x_1}=\rho\frac{\partial^2 u_1}{\partial t^2},\\
&(c_{55}+c_{31})\frac{\partial^2 u_1}{\partial x_3 \partial x_1}+c_{55}\frac{\partial^2 u_3}{\partial {x_1}^2}+c_{33}\frac{\partial^2 u_3}{\partial {x_3}^2}+(e_{15}\frac{\partial^2 \phi}{\partial {x_1}^2}+e_{33}\frac{\partial^2 \phi}{\partial {x_3}^2})=\rho\frac{\partial^2 u_3}{\partial t^2},\\
&(e_{15}+e_{31})\frac{\partial u_1}{\partial x_3 \partial x_1}+(e_{15}\frac{\partial u_3}{\partial {x_1}^2}+e_{33}\frac{\partial u_3}{\partial {x_3}^2})-(\varepsilon_{11}\frac{\partial \phi}{\partial {x_1}^2}+\varepsilon_{33}\frac{\partial \phi}{\partial {x_3}^2})=0.
\end{aligned}
\tag{6}
$$

For time-harmonic motion, the solutions of Eq. (6) take the following form:

$$
\begin{aligned}
u_1 &= B_1 e^{kbz} e^{ik(x-ct)},\\
u_3 &= B_2 e^{kbz} e^{ik(x-ct)},\\
\varphi &= B_3 e^{kbz} e^{ik(x-ct)},
\end{aligned}
\tag{7}
$$

where $k$ is the wave number，$c$ is the phase velocity，$B_i (i=1,2,3)$ are amplitudes, $b$ is an unknown constant to be determined.

Substituting Eq. (7) into Eq. (6) yields

$$
\begin{aligned}
&(-k^2c_{11}+k^2b^2c_{55}+k^2c^2\rho)A+ik^2b(c_{13}+c_{55})B+ik^2b(e_{15}+e_{31})C=0,\\
&ik^2b(c_{55}+c_{31})A+(k^2c^2\rho-k^2c_{55}+k^2b^2c_{33})B+(-k^2e_{15}+k^2b^2e_{33})C=0,\\
&ik^2b(e_{15}+e_{31})A+(-k^2e_{15}+k^2b^2e_{33})B+(k^2\varepsilon_{11}-k^2b^2\varepsilon_{33})C=0.
\end{aligned}
\tag{8}
$$

Eq. (8) can be written as

$$
\mathbf{M}=\begin{vmatrix}
c^2\rho+b^2c_{55}-c_{11} & ib(c_{13}+c_{55}) & ib(e_{15}+e_{31})\\
ib(c_{55}+c_{31}) & c^2\rho+b^2c_{33}-c_{55} & b^2e_{33}-e_{15}\\
ib(e_{15}+e_{31}) & b^2e_{33}-e_{15} & \varepsilon_{11}-b^2\varepsilon_{33}
\end{vmatrix}.
\tag{9}
$$

To obtain a non-trivial solution of the Eq. (9), the determinant of the coefficient matrix **M** must be equal to zero, and the obtained equation is

$$
a_6\, b^6+a_4\, b^4+a_2\, b^2+a_0=0,
\tag{10}
$$

where the detailed expressions of $a_i$ ($i$=0, 2, 4, 6) are respectively

$$
\begin{aligned}
&a_0=(c_{11}-c^2\rho)(e_{15}^2+\varepsilon_{11}(c_{55}-c^2\rho)),\\
&a_2=c_{13}e_{15}^2+c_{31}e_{15}^2+c_{13}e_{15}e_{31}+c_{31}e_{15}e_{31}-c_{55}e_{31}^2-2c_{11}e_{15}e_{33}+c^2e_{15}^2\rho+2c^2e_{15}e_{31}\rho\\
&+c^2e_{31}^2\rho+2c^2e_{15}e_{33}\rho+\varepsilon_{33}(c_{11}-c^2\rho)(-c_{55}+c^2\rho)\\
&+\varepsilon_{11}(-c_{11}c_{33}+c_{31}c_{55}+c_{13}(c_{31}+c_{55})+c^2c_{33}\rho+c^2c_{55}\rho),\\
&a_4=-(-\varepsilon_{11}c_{33}c_{55}-c_{33}e_{15}^2-2c_{33}e_{15}e_{31}-c_{33}e_{31}^2+c_{13}e_{15}e_{33}+c_{31}e_{15}e_{33}+c_{13}e_{31}e_{33}+c_{31}e_{31}e_{33}\\
&+2c_{55}e_{31}e_{33}-c_{11}e_{33}^2+c^2e_{33}^2\rho+\varepsilon_{33}(-c_{11}c_{33}+c_{31}c_{55}+c_{13}(c_{31}+c_{55})+c^2c_{33}\rho+c^2c_{55}\rho)),\\
&a_6=-c_{55}(\varepsilon_{33}c_{33}+{e_{33}}^2).
\end{aligned}
\tag{11}
$$

Eq. (10) is solved to get $b_{1(n)}-b_{6(n)}$. In a piezoelectric substrate, when $x_3$ tends to infinity, namely $x_3 \to +\infty$,

$$u_3^{(m)} = 0,\ \phi^{(m)} = 0. \tag{12}$$

Three negative roots are taken according to Eq. (12). For a given value of $c$, by solving the equation, each value of $b$ represents an independent solution of Eqs. (1-3). Therefore, the displacement and potential in the piezoelectric substrate should be further represented in a linear combination, namely

$$\begin{aligned} u_1 &= \left[\sum_{n=1}^{3} f_{1n} B_{3n} \exp(kb_n z)\right] \exp[ik(x-ct)], \\ u_3 &= \left[\sum_{n=1}^{3} f_{2n} B_{3n} \exp(kb_n z)\right] \exp[ik(x-ct)], \\ \varphi &= \left[\sum_{n=1}^{3} B_{3n} \exp(kb_n z)\right] \exp[ik(x-ct)], \end{aligned} \tag{13}$$

where $f_{1n}, f_{2n}$ are the amplitude ratios that can be defined from Eq. (9), namely

$$\begin{aligned} f_{1n} &= \frac{M_{12}M_{23} - M_{13}M_{22}}{M_{11}M_{22} - M_{12}M_{21}}, \\ f_{2n} &= \frac{M_{13}M_{21} - M_{11}M_{23}}{M_{11}M_{22} - M_{12}M_{21}}. \end{aligned} \tag{14}$$

Equations of motion and stresses of viscous liquid in the $x$-$z$ plane are presented [47] as follows:

$$\begin{aligned} \rho^L \frac{\partial^2 \varphi^L}{\partial t^2} &= \lambda^L \nabla^2 \varphi^L + \frac{4}{3}\eta^L \frac{\partial}{\partial t} \nabla^2 \varphi^L, \\ \rho^L \frac{\partial^2 \psi^L}{\partial t^2} &= \eta^L \frac{\partial}{\partial t} \nabla^2 \psi^L, \end{aligned} \tag{15}$$

and

$$\begin{aligned} \sigma_{zz}^L &= \left[\lambda^L - \frac{2}{3}\eta^L \frac{\partial}{\partial t}\right]\left[\varphi_{,xx}^L + \varphi_{,zz}^L\right] + 2\eta^L \frac{\partial}{\partial t}\left[\psi_{,xz}^L + \varphi_{,zz}^L\right], \\ \sigma_{xz}^L &= \eta^L \frac{\partial}{\partial t}\left[2\varphi_{,xz}^L + \psi_{,xx}^L - \psi_{,zz}^L\right], \end{aligned} \tag{16}$$

where $\lambda^L$, $\rho^L$ and $\eta^L$ are bulk modulus, density, and viscosity of liquid, respectively. The potentials of longitudinal waves and transverse waves are given respectively by

$$u_1^L = \varphi_{,1}^L - \psi_{,3}^L,\ u_3^L = \varphi_{,3}^L + \psi_{,1}^L. \tag{17}$$

The solutions of Eq. (13) are with the following form:

$$\varphi^L(x,z,t) = A_1\exp(n_1x_3)\exp[ik(x_1-ct)] + A_2\exp(n_2x_3)\exp[ik(x_1-ct)],$$
$$\psi^L(x,z,t) = A_3\exp(n_3x_3)\exp[ik(x_1-ct)] + A_4\exp(n_4x_3)\exp[ik(x_1-ct)], \quad (18)$$

where $A_1$, $A_2$, $A_3$ and $A_4$ are constants to be determined, $n_1$, $n_2$, $n_3$ and $n_4$ are respectively

$$n_1 = \sqrt{k^2 - \frac{k^2c^2\rho^L}{\lambda^L - \frac{4}{3}ikc\eta^L}}, n_2 = -\sqrt{k^2 - \frac{k^2c^2\rho^L}{\lambda^L - \frac{4}{3}ikc\eta^L}},$$
$$n_3 = \sqrt{k^2 - \frac{ikc\rho^L}{\eta^L}}, n_4 = -\sqrt{k^2 - \frac{ikc\rho^L}{\eta^L}}. \quad (19)$$

Boundary and continuous conditions are as the following. At $x_3 = -h$, the boundary condition of fluid is:

$$\sigma^L_{13(x_1,-h)} = 0,\ \sigma^L_{33(x_1,-h)} = 0. \quad (20)$$

At $x_3 = 0$, continuous conditions at the interface between the fluid and piezoelectric substrate are

$$u^L_{1(x_1,0)} = u_{1(x_1,0)},\ u^L_{3(x_1,0)} = u_{3(x_1,0)},\ \sigma^L_{33(x_1,0)} = \sigma_{33(x_1,0)},\ \sigma^L_{13(x_1,0)} = \sigma_{13(x_1,0)},\ D_{3(x_1,0)} = 0. \quad (21)$$

Substitute Eq. (3), (13), (16) and (17) to the above boundary and continuous conditions yields

$$[-(\lambda^L + \frac{2}{3}\eta^L ikc)k^2 + (\lambda^L + \frac{2}{3}\eta^L ikc - 2ikc\eta^L)n_1^{\,2}]\exp(n_1x_3)A_1$$
$$+[-(\lambda^L + \frac{2}{3}\eta^L ikc)k^2 + (\lambda^L + \frac{2}{3}\eta^L ikc - 2ikc\eta^L)n_2^{\,2}]\exp(n_2x_3)A_2 \quad (22.a)$$
$$+k^2c2\eta^L n_3\exp(n_3x_3)A_3 + k^2c2\eta^L n_4\exp(n_4x_3)A_4 = 0,$$

$$2\eta^L k^2cn_1\exp(n_1x_3)A_1 + 2\eta^L k^2cn_2\exp(n_2x_3)A_2 +$$
$$(\eta^L ick^3 + \eta^L ikcn_3^{\,2})\exp(n_3x_3)A_3 + (\eta^L ick^3 + \eta^L ikcn_4^{\,2})\exp(n_4x_3)A_4 = 0, \quad (22.b)$$

$$\sum_{n=1}^{3} f_{1n}B_{3n} - ikA_1 - ikA_2 + n_3A_3 + n_4A_4 = 0, \quad (22.c)$$

$$\sum_{n=1}^{3} f_{2n}B_{3n} - n_1A_1 - n_2A_2 - ikA_3 - ikA_4 = 0, \quad (22.d)$$

$$[-(\lambda^L + \frac{2}{3}\eta^L ikc)k^2 + (\lambda^L + \frac{2}{3}\eta^L ikc - 2ikc\eta^L)n_1^{\,2}]A_1$$
$$+[-(\lambda^L + \frac{2}{3}\eta^L ikc)k^2 + (\lambda^L + \frac{2}{3}\eta^L ikc - 2ikc\eta^L)n_2^{\,2}]A_2 \quad (22.e)$$
$$+k^2c2\eta^L n_3A_3 + k^2c2\eta^L n_4A_4 - k\sum_{n=1}^{3} B_{3n}(if_{1n}c_{31} + b_nc_{33}f_{2n} + b_ne_{33}) = 0,$$

$$k\sum_{n=1}^{3} B_{3n}((c_{55}(f_{1n}b_n + if_{2n}) + ie_{51}) - 2\eta^L k^2 cn_1 A_1 - 2\eta^L k^2 cn_2 A_2 \\ -(\eta^L ick^3 + \eta^L ikcn_3^{\,2})A_3 - (\eta^L ick^3 + \eta^L ikcn_4^{\,2})A_4 = 0, \tag{22.f}$$

$$\sum_{n=1}^{3} B_{3n}(if_{1n}e_{31} + b_n e_{33} f_{2n} - b_n \varepsilon_{33}) = 0. \tag{22.g}$$

For nontrivial solutions, the determinant of the coefficient matrix of Eq. (22.a-g) has to vanish, which yields an equation that determines dispersion curves (phase velocity $c$ vs. wave number $k$). The parameters of the viscous liquid layer are set as: $\lambda^L = 22\times10^8\,\text{N/m}^2, \rho^L = 1000\sim1500\,\text{kg/m}^3, \eta^L = 0.2\sim1.2\,\text{Pa•s}$ and $h = 10\sim20\ \mu\text{m}$. The material of the piezoelectric substrate is set as Lithium Niobate ($LiNbO_3$) single crystal, the parameters of which are listed in Table 1 [48,49].

**Table 1.** Material parameters of $LiNbO_3$ single crystal

| Properties | value | Properties | value |
|---|---|---|---|
| $c_{11}$ (GPa) | 203 | $e_{31}$(C/m$^2$) | 1.6968 |
| $c_{33}$ (GPa) | 8.53 | $e_{33}$(C/m$^2$) | 2.3227 |
| $c_{13}$ (GPa) | 57.93 | $\varepsilon_{11}$($10^{-9}$ C/Vm) | 44 |
| $c_{55}$ (GPa) | 56.95 | $\varepsilon_{33}$($10^{-9}$ C/Vm) | 34.685 |
| $e_{15}$ (C/m$^2$) | 4.4548 | $\rho$ (kg/m$^3$) | 4700 |

The influences of thickness, density and viscosity of the liquid-layer loading on the dispersion curves of Rayleigh SAW are determined，shown in Fig.2. The results show that the thickness, density and viscosity of the liquid layer all can lead to changes of dispersion characteristics. The phase velocity of SAW decreases with the increase of the thickness and density of the liquid-layer loading (Fig.2(a-b)), and increases with the increase of viscosity (Fig.2(c)).

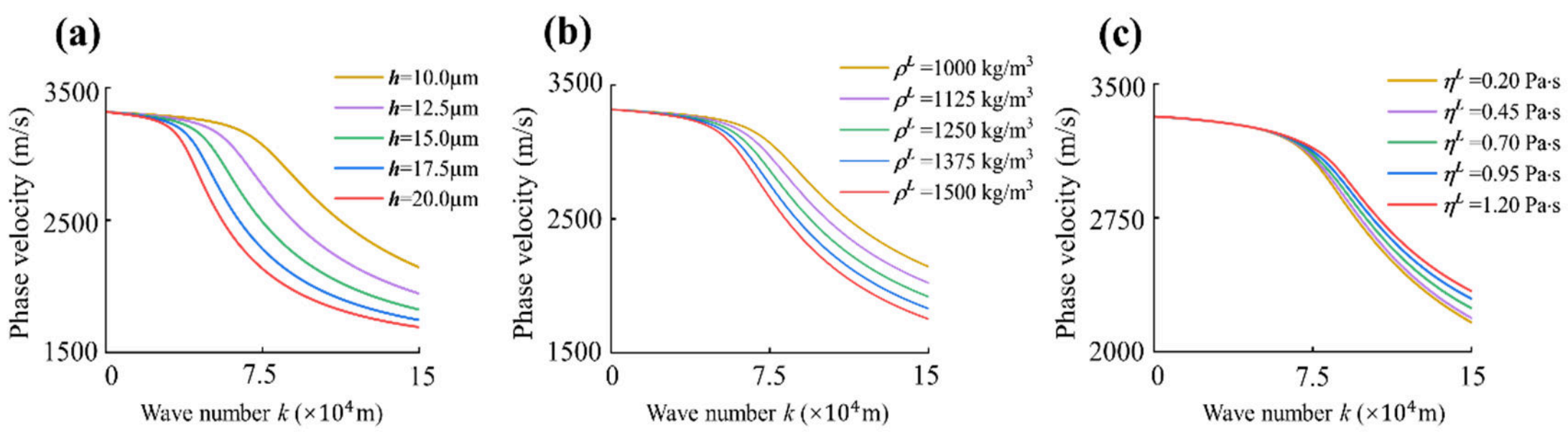

**Fig. 2.** Dispersion curves ($c$ vs. $k$) of Rayleigh SAW in $LiNbO_3$ single crystal under a liquid-layer loading, (a) with different liquid-layer thicknesses, (b) with different liquid densities, (c) with different liquid viscosities.

For periodic phononic crystals or metamaterials, the propagation velocity of waves influences Bragg scattering conditions obviously. The group velocity (a function of phase velocity) of the wave is proportional to the slope of the dispersion curve. Thus, changes in surface liquid-layer loadings enable the tuning and manipulation of the group velocity of waves， which influences the position and width of the band gap, and the location and number of Dirac cones in dispersion curves of metamaterials. Revealing of influence laws of the surface liquid-layer loading on the wave phase velocity can be used to guide the realization of specific energy bands. This brings a new idea for designing a liquid sensor based on topological phononic crystals.

The model of SAW topological phononic crystals covered by a liquid-layer loading is constructed, shown in **Fig. 3**. The phononic crystal consists of a *y*-cut $LiNbO_3$ substrate and two sets of micro resonant pillars embedded within the substrate, labeled as A and B. Two kinds of cylindrical pillars have the same height ($h = 20\ \mu m$), side angle $(\theta = 96^{\circ})$ and different radii ($r_A$, $r_B$). The pillars are made of lithium (the density is $\rho = 534\ kg/m^3$, Young's modulus is $E = 4.9\times10^9\,Pa$, Poisson's ratios is 0.381). The surface of the structure is covered by a liquid-layer loading.

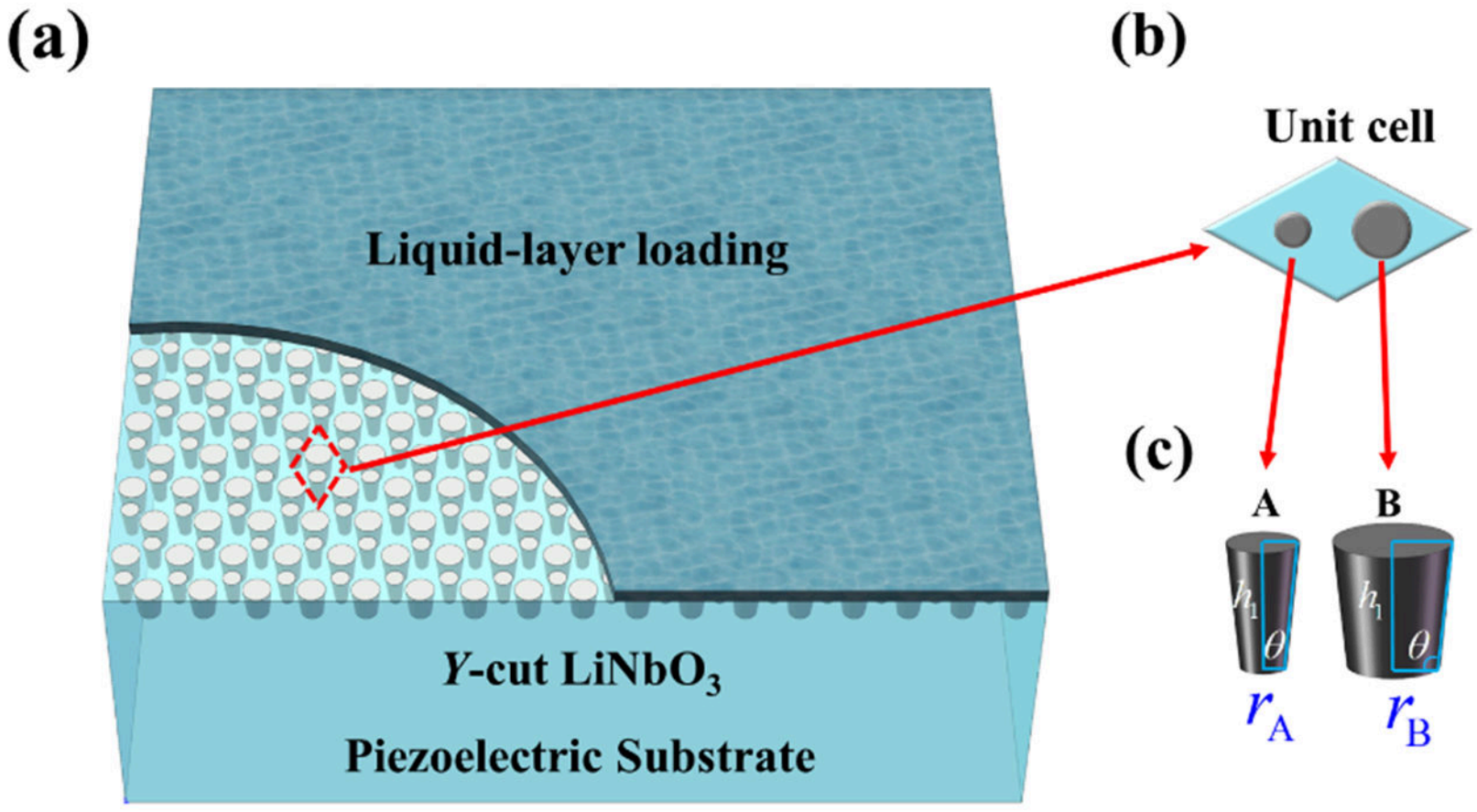

**Fig. 3.** Schematic diagram of the SAW topological phononic-crystal device under a surface liquid-layer loading. (a) The phononic crystal consists of a *y*-cut LiNbO3 substrate and two sets of micro resonant pillars embedded within the substrate, labeled as A and B. (b) The unit cell of the phononic crystal with a lattice constant $a = 20\sqrt{3}\ \mu m$. (c) Two resonant pillars with different radii ($r_A = 3.9\ \mu m$, $r_B = 7\ \mu m$) and the same side angle ($\theta$=96$^{\circ}$).

Influence laws of the surface liquid-layer loading on the wave phase velocity shown in Fig.2 are used to guide the parameter tuning of the liquid-layer loading. The dispersion characteristics of the unit cell of the phononic crystal covered with a liquid-layer loading are numerically simulated by using the finite element method. Soft boundary conditions are applied at the interface between the liquid loading and the air, namely

$$p = 0, \tag{23}$$

where $p$ is the pressure in the fluid. The generation of acoustic waves in the air is neglected. At the interface between the liquid layer loading and the solid, continuous conditions are applied, namely

$$\frac{1}{\rho_f}\frac{\partial \rho}{\partial n} = \omega^2 u_n,\ \ \sigma_{ij} n_{ij} = p n_j, \tag{14}$$

where $\sigma_{\mathrm{ij}}$ is the Cauchy stress tensor, $u_{\mathrm{n}}$ is the normal displacement of the solid boundary and **n** is the normal unit vector inside the liquid loading. Considering the pressure wave equations in the fluid and the elastic dynamic equations in the solid, which couple through continuous conditions, the discrete form of the acoustic-elastic equations can be expressed as follows [50–52]:

$$\begin{pmatrix} \mathbf{K}_s & \mathbf{S}_{fs}^T \\ \mathbf{0} & \mathbf{K}_f \end{pmatrix}\begin{pmatrix} \mathbf{u} \\ \mathbf{p} \end{pmatrix} - \omega^2 \begin{pmatrix} \mathbf{M}_s & \mathbf{0} \\ -\mathbf{S}_{fs} & \mathbf{M}_f \end{pmatrix}\begin{pmatrix} \mathbf{u} \\ \mathbf{p} \end{pmatrix} = \begin{pmatrix} \mathbf{F} \\ 0 \end{pmatrix}, \tag{25}$$

where $\mathbf{u}$ and $\mathbf{p}$ represent the displacement and pressure at mesh nodes, respectively, and $\mathbf{F}$ is nodal force. $\mathbf{K}_s$ and $\mathbf{K}_f$ represent the stiffness matrices of the solid and fluid, respectively; $\mathbf{M}_s$ and $\mathbf{M}_f$ are the mass matrices of the solid and fluid, respectively; $\mathbf{S}_{fs}$ represents the fluid-solid coupling matrix. According to Eq. (25), the pressure in fluids and displacement in solids can be obtained.

The dispersion band structure of the unit cell can be obtained by solving the eigenvalue problem. To check the influences of the liquid-layer loading on the dispersion band structure of phononic crystals, the differences between the band structures of phononic crystals with and without liquid-layer loadings are compared in **Appendix A** ( for the unit cell, $r_A = 3.9\ \mu\text{m}, r_B = 7\ \mu\text{m}$). It is noted that compared to the band structure without a liquid-layer loading, the addition of the liquid layer loading introduces several extra bands obviously. When the liquid layer is added on the surface, the dispersion of bands is significantly affected. Owing to the fluid-solid boundary condition and local-resonance mechanism, the pressure waves appear in the liquid, which becomes e available modes at the particular excitation frequencies.

Fig.4 shows the dispersion band structures and topological phase transitions of different types of phononic crystals under a liquid-layer loading, including Type-A unit cell ( $r_A = 3.9\ \mu m, r_B = 7\ \mu m$ ), Type-B unit cell ($r_A = r_B = 7\ \mu m$) and Type-C unit cell ($r_A = 7\ \mu m, r_B = 3.9\ \mu m$) . Because of the presence of the liquid-layer loading, energy bands resulting from fluid-solid couplings appear within the sound lines of the SAW device. Fig.4(b) shows that when the radii of two kinds of scatterers are equal (Type-B unit cell), the Dirac cones at M point within the first Brillouin zone are closed. However, when a difference in radii is introduced (Type-A and Type-C unit cells) shown in Fig.4 (a) and (c) respectively, the spatial inversion symmetry is broken, resulting in a significant energy disparity between the two valleys. Consequently, energy flow between the valleys is impeded, giving rise to a band gap, which is the valley Hall effect. The inset in Fig. 4 (a-c) show displacement fields and energy flows at M point, where the displacement distributions are represented by the color plots, and the directions of the energy flows are indicated by the red arrows. For the Type-A and Type-C unit cell, energy flows are distributed in opposite directions. This is an important characteristic of the valley modes, and it indicates that a topological phase transition has occurred. Fig. 4 (a-c) only present a part of the complete band structure. The full band structures can be seen in **Appendix B**. We also verify that surface acoustic wave mode exists indeed in this structure (**Appendix C**). Fig.4(d) shows the curve of band-gap size vs. the radius difference of the scatterers ( $\Delta r = r_A - r_B$ ), it can be seen that the band gap of the SAW topological phononic crystals can be tuned by changing the radius difference of the scatterers. As the absolute value of $\Delta r$ increases from zero, the degree of the symmetry breaking becomes higher and the width of the band gap becomes larger accordingly.

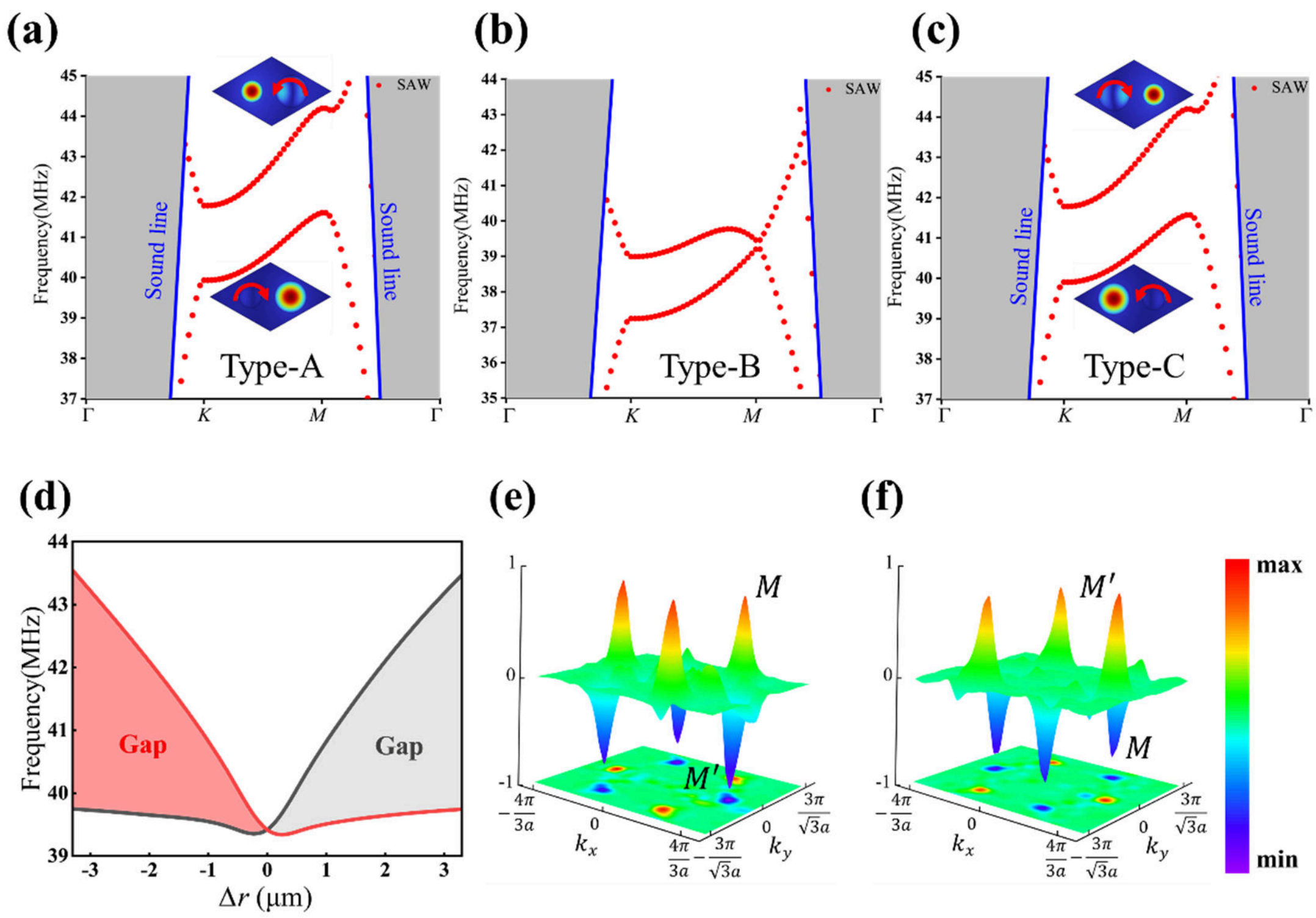

**Fig. 4.** The dispersion band structures and topological phase transitions of different types of phononic crystals under a liquid-layer loading. (a) The band structure of the Type-A unit cell $(r_A = 3.9\ \mu\text{m}, r_B = 7\ \mu\text{m})$. (b) The band structure of the B-type unit cell $(r_A = r_B = 7\ \mu\text{m})$. (c) The band structure of the Type-C unit cell $(r_A = 7\ \mu\text{m}, r_B = 3.9\ \mu\text{m})$. The Type-A and Type-C unit cells exhibit the opening of Dirac cones at the M and M' points within the first Brillouin zone. The insets show energy flow distributions of phononic crystals with different topological phases. (d) Band gap vs. the radius difference of the scatterers ( $\Delta r = r_A - r_B$ ). (e-f) The Berry curvatures around M and M' points, respectively.

The valley Chern number and Berry curvature are two important parameters for characterizing the topological properties of dispersion bands. To further verify the generation of the topological phase transition from a physical perspective, quantitative Berry curvature and topological valley Chern number are calculated by using the discrete method. Compared to the qualitative approach using k·p perturbation theory, the discrete numerical methods can capture changes in valley Chern numbers and Berry curvature more accurately [53,54]. The topologically nontrivial properties of the SAW system are indicated by a nonzero valley Chern number, which is the integral of Berry curvature around the M and M' valleys, namely

$$C_{\mathrm{V}}^{(n)}=\frac{1}{2\pi}\int\mathbf{\Omega}(\mathbf{k})d^2\mathbf{k},$$

where $\mathbf{\Omega}(\mathbf{k})$ is the Berry curvature, namely

$$\mathbf{\Omega}(\mathbf{k})=\mathrm{i}\nabla_{\mathrm{k}}\times\langle\mathbf{u}(\mathbf{k})|\nabla_{k}|\mathbf{u}(\mathbf{k})\rangle,$$

where $\mathbf{u}(\mathbf{k})$ represents the displacement field of the eigen mode computed by using the COMSOL Multiphysics. The detailed calculation process of the Berry curvature can be found in **Appendix D**, the distributions of Berry curvature are shown in Fig. 4 (e-f). The Berry curvature predominantly resides near the M and M' valleys. Moreover, there is a significant sign reversal observed when a switch of unit-cell type occurs between A-type and C-type, confirming the occurrence of a topological phase transition. After integrating the Berry curvature to obtain the valley Chern number, it is typically expected to be either 1/2 or -1/2. However, the numerical calculation in this work revealed a Chern number is smaller than 1/2, which is attributed to the strong breaking of spatial inversion symmetry [55,56].

The band structures and Berry curvatures corresponding to unit cells under loadings with different liquid-layer thickness are calculated. The liquid used in the calculation of the band structure is water. The results are shown in 错误!未找到引用源。5. From 错误!未找到引用源。5 (a)-(c), it can be seen that as the liquid-layer thickness increases from 8 to 13 μm, the position of the bandgap decreases from 43 MHz to 36 MHz. Furthermore, with an increasing liquid-layer thickness, the intensity of the Berry curvature increases sharply, and Chern number increases obviously, indicating that the thicknesses of the liquid loading have a significant influence on the degree of the spatial inversion symmetry breaking and topological indices. This result provides a new, effective and convenient approach (changing the thickness of the liquid-layer) for tuning topological Chern number of topological SAW devices.

In addition, in **Appendix D**, the variations of Berry curvature are calculated under different symmetry-breaking degrees through tuning the radius difference of scatterers ($\Delta r=r_A-r_B$).

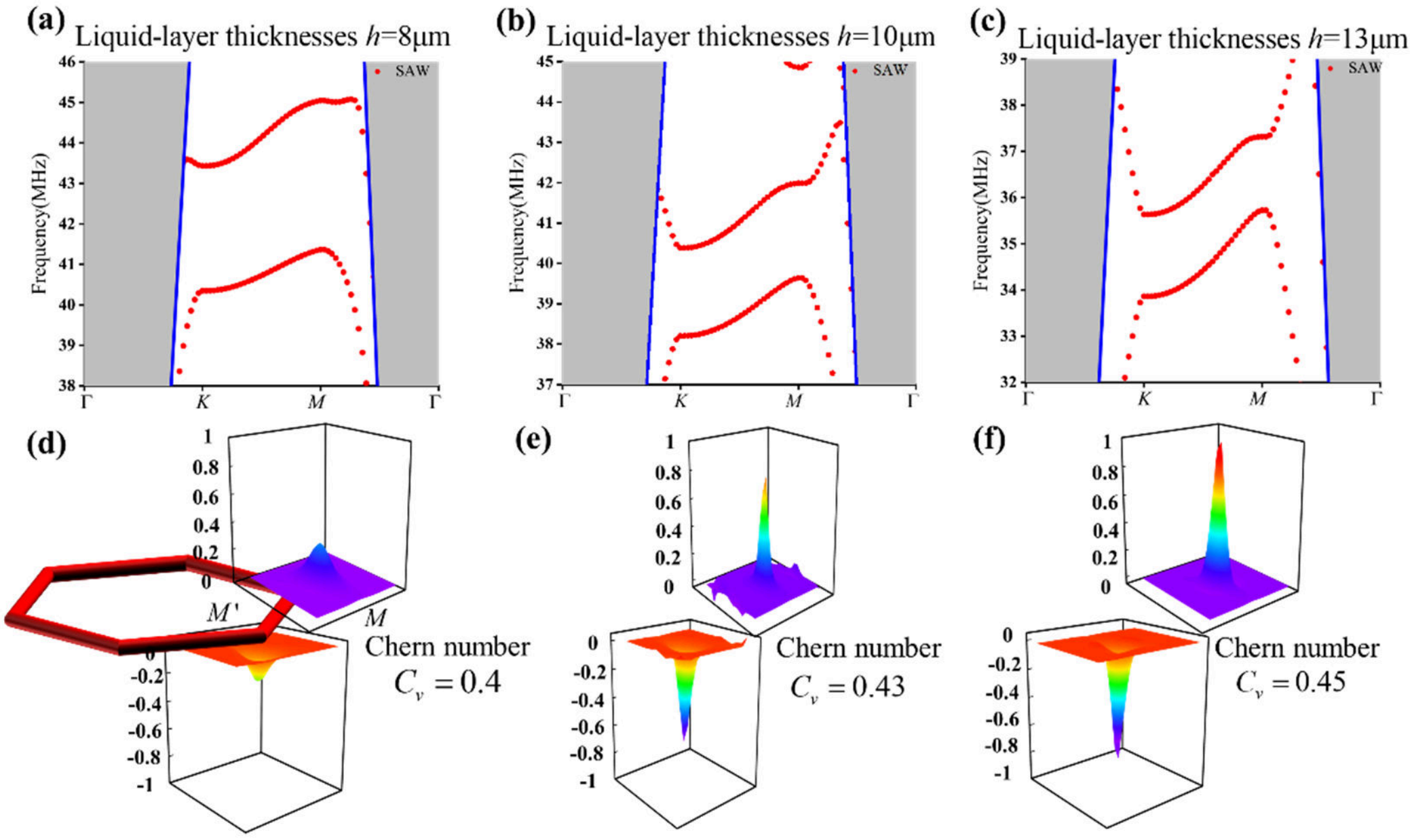

**Fig. 5.** The band structures and topological indices (Berry curvature and Chern number) corresponding to unit cells under loadings with different liquid-layer thicknesses. (a)-(c). The bandgaps in the band structure of the unit cells with the increasing liquid-layer thicknesses (8, 10, 13 μm). As the liquid-layer thickness increases, the bandgap gradually decreases. (d)-(f). Berry curvatures and Chern numbers corresponding to (a)-(c). As the liquid-layer thickness increases, the intensity of the Berry curvature around the M and M' points increases sharply and the Chern number increases obviously.

# 3. Topological resonance behaviors of SAW under a surface liquid-layer loading

In this section, the couplings of the topological waveguide and resonant cavity under a surface liquid-layer loading are utilized to create topological resonance peaks with a high Q-factor.

First, topological transmissions of SAW under a liquid-layer loading are realized, and influences of liquid-layer loadings on those are revealed, the details are shown in **Appendix E**. Then, a valley insulator with a defect area is used to establish a resonant cavity. As shown in Fig. 6 (a), in the area within the dotted box, type-C phononic crystals are replaced by type-B ones. As a result, a defect (area within the dotted box) of the topological insulator emerges, effectively forming a resonant cavity. The resonance frequency points of the resonant cavity within the band gap range of 39-43 MHz are noted by red points in Fig. 6 (c). The displacement mode of the resonant cavity is depicted in Fig. 6 (b), it is shown that the energy is mainly concentrated inside the cavity.

The coupling system of the straight topological waveguide and the resonant cavity is constructed, shown in Fig. 7 (a). A topological straight waveguide is formed along the interface between the type-A and type-C phononic crystals, and a resonant cavity (type-B phononic crystals) is set in type-C phononic crystals. The transmission spectrum of the coupling system is shown in Fig. 7(b). It is shown that there emerges an obvious resonance peak, which results from the coupling of the resonant cavity and the straight waveguide. The displacement distributions of the modes at the peak (red) and the allowed mode (blue) are depicted in Fig. 7(c) and (d), respectively. It can be seen that for the mode at the peak (resonance mode), energy is almost entirely concentrated within the resonant cavity, while for the allowed states, SAW can transport through the waveguide mostly. The resonance peak formed by interference between the straight wave guide and the resonant cavity under a surface liquid-layer loading provides favorable conditions for realizing liquid-phase sensing.

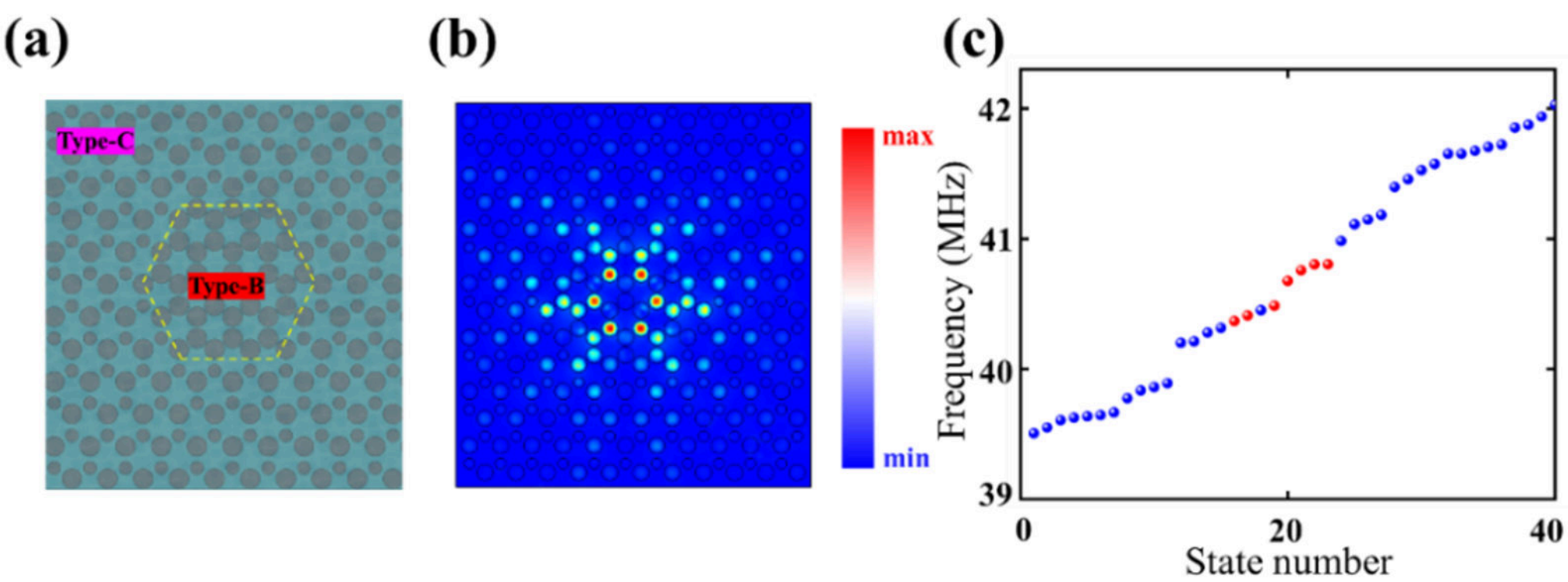

**Fig. 6.** (a) Schematic diagram of the resonant cavity. (b) The displacement distribution of the eigenmode of the resonant cavity at 40.676 MHz. (c) The eigen frequencies of the resonant cavity within 39-42 MHz. The resonance modes are represented in red, while the bulk modes are represented in blue.

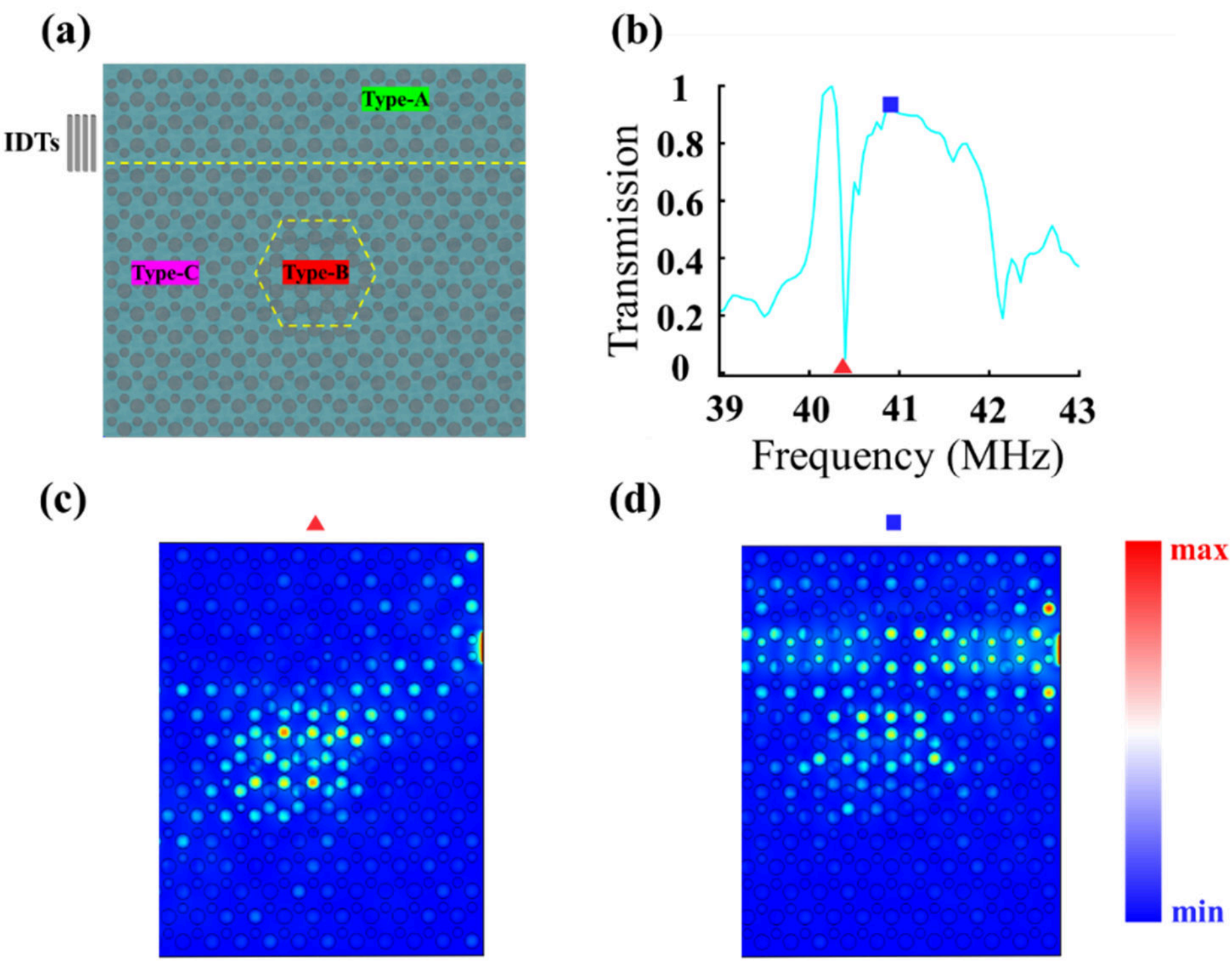

**Fig. 7.** The generating of the resonance peak of the waveguide-resonator coupling system under a liquid-layer loading. (a) Schematic diagram of the coupling system including a topological interface waveguide and a resonant cavity under a surface liquid-layer loading. (b) Normalized transmission spectrum of the coupling system, the resonance (peak) mode is indicated by the red triangle and the allowed mode is indicated by the blue square. (c) and (d) The displacement distributions at the resonance-mode frequency and allowed-mode frequency, respectively.

The robustness of transmission spectrums of the waveguide-resonator coupling system is verified through numerical simulations. Fig. 8 (a) shows the coupling system without a defects or disorder. Fig. 8 (b) shows the coupling systems with defects on the straight waveguide path, namely two pillars are deleted. Besides, a disorder is added on the straight waveguide path, namely, the diameter of two thinner pillars is increased from 3.9 to 5.2 mm, shown in Fig. 8 (c). The normalized transmission spectrums of the three cases are plotted in Fig. 8 (d). When the defect or disorder is introduced in the SAW transmission path, the frequency of the resonance peak almost does not change. It is indicated that the topology of the system ensures that the resonance peak is not influenced obviously by the appearance of the defect or disorder.

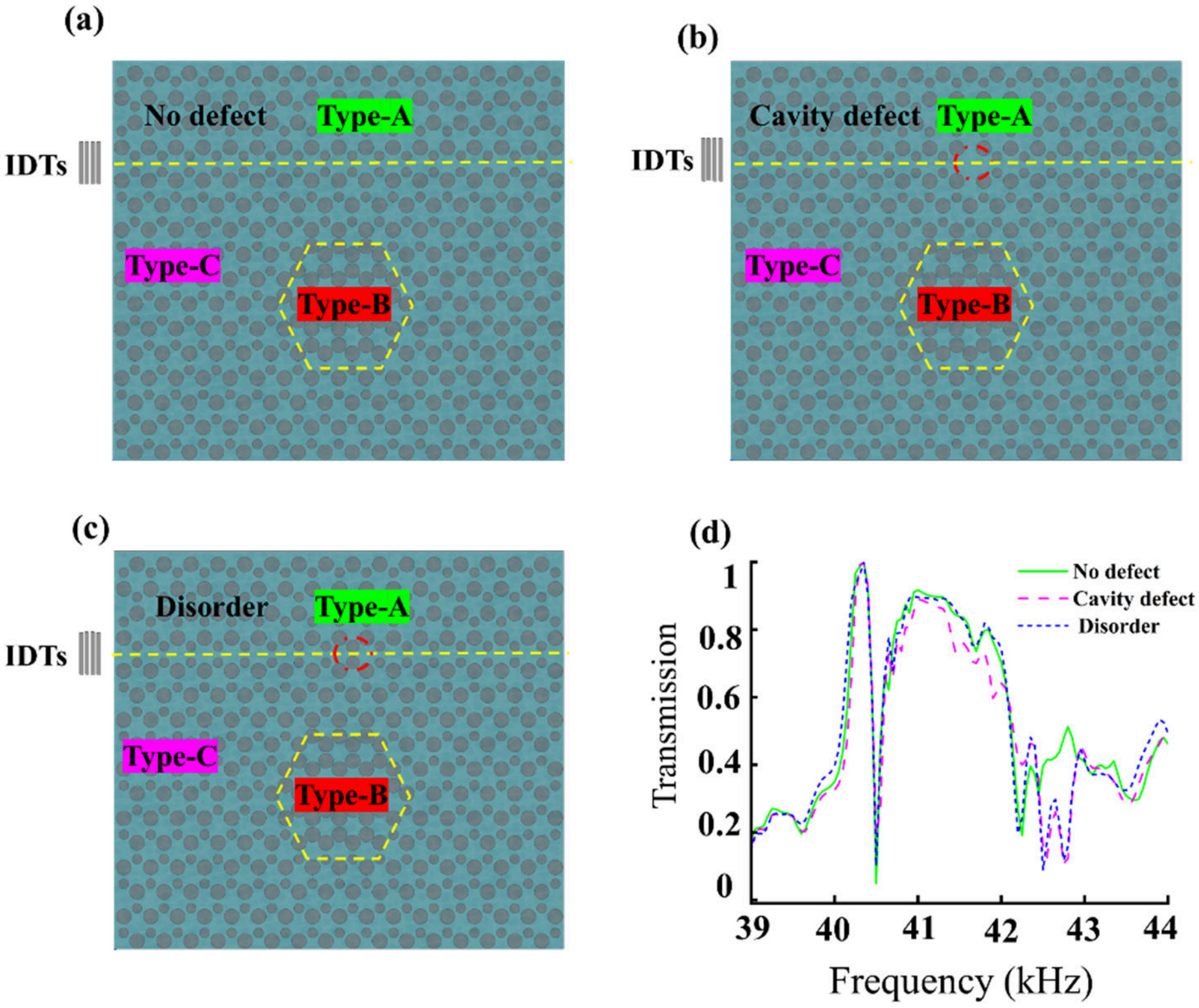

**Fig. 8.** Analysis of the robustness of the topological resonance generated by couplings between a straight waveguide and a resonant cavity under a liquid-layer loading. (a)-(c) The coupling systems under three different conditions: with no defect, with cavity defects, and with disorder, respectively. (d) The normalized transmission spectrums of the three cases.

Then, influences of structural parameters of the system on characteristics of the topological resonance are analyzed. Firstly, the radius difference of pillars A and B in the unit cell is changed. Here, three cases ($r_A = 3.9$ μm, $r_B = 7.2$ μm; $r_A = 4.6$ μm, $r_B = 7.2$ μm; $r_A = 5.2$ μm, $r_B = 7.2$ μm) are considered. The resonance peaks generated by the coupling system are shown in Fig. 9. The results show that the Q-factor has a trend of decreasing with the decreasing radius difference, which is due to the weakening of the degree of the spatial-inversion-symmetry breaking. Besides, the shifts of the resonance frequency occurs (the resonance frequencies for cases in Fig. 9.(b-d) are 40.4 MHz, 40.3 MHz and 40.25 MHz, respectively).

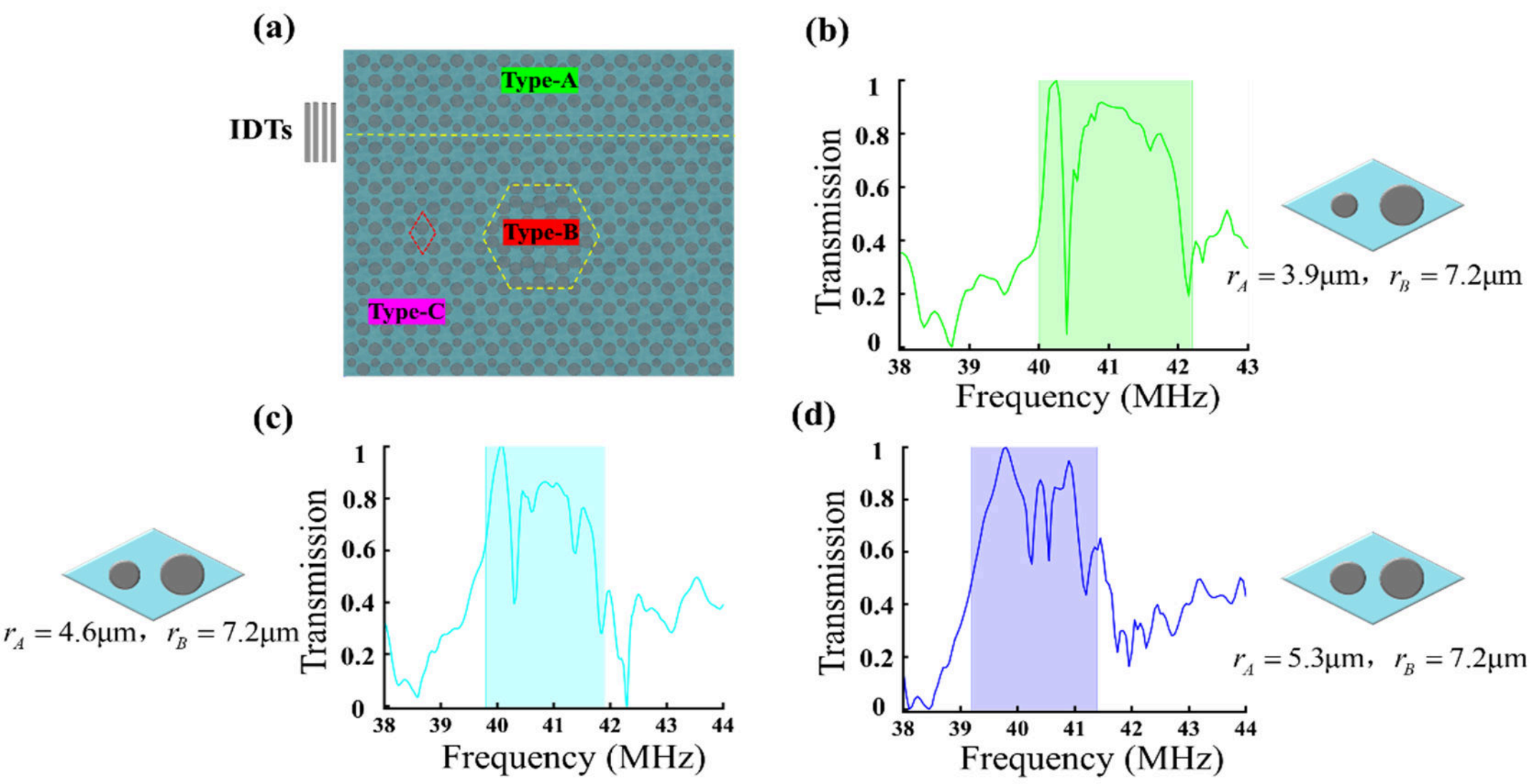

**Fig. 9.** The influences of the degree of the spatial-inversion-symmetry breaking (radius differences of pillars in the unit cell) on characteristics of the topological resonance. (a) A coupling system including a straight waveguide and a resonant cavity under a liquid-layer loading. (b)-(d) Normalized transmission spectrums of the coupling system with different radius differences of pillars in the unit cell.

Besides, influences of the liquid-layer thickness on characteristics of the topological resonance are checked. The liquid-layer thickness is set to $h = 10\ \mu m$, $h = 11\ \mu m$, and $h = 12\ \mu m$, respectively, and the resonance frequencies and Q-factors are compared, shown in Fig. 10. The results show that as the liquid-layer thickness increases, the frequency of the topological resonance changes obviously (the resonance frequencies for cases in Fig. 10 (b-d) are 40.4 MHz, 38.8 MHz and 37.35 MHz, respectively). Additionally, as the thickness of the surface liquid- layer increases, there is a slight decrease in the Q-factor.

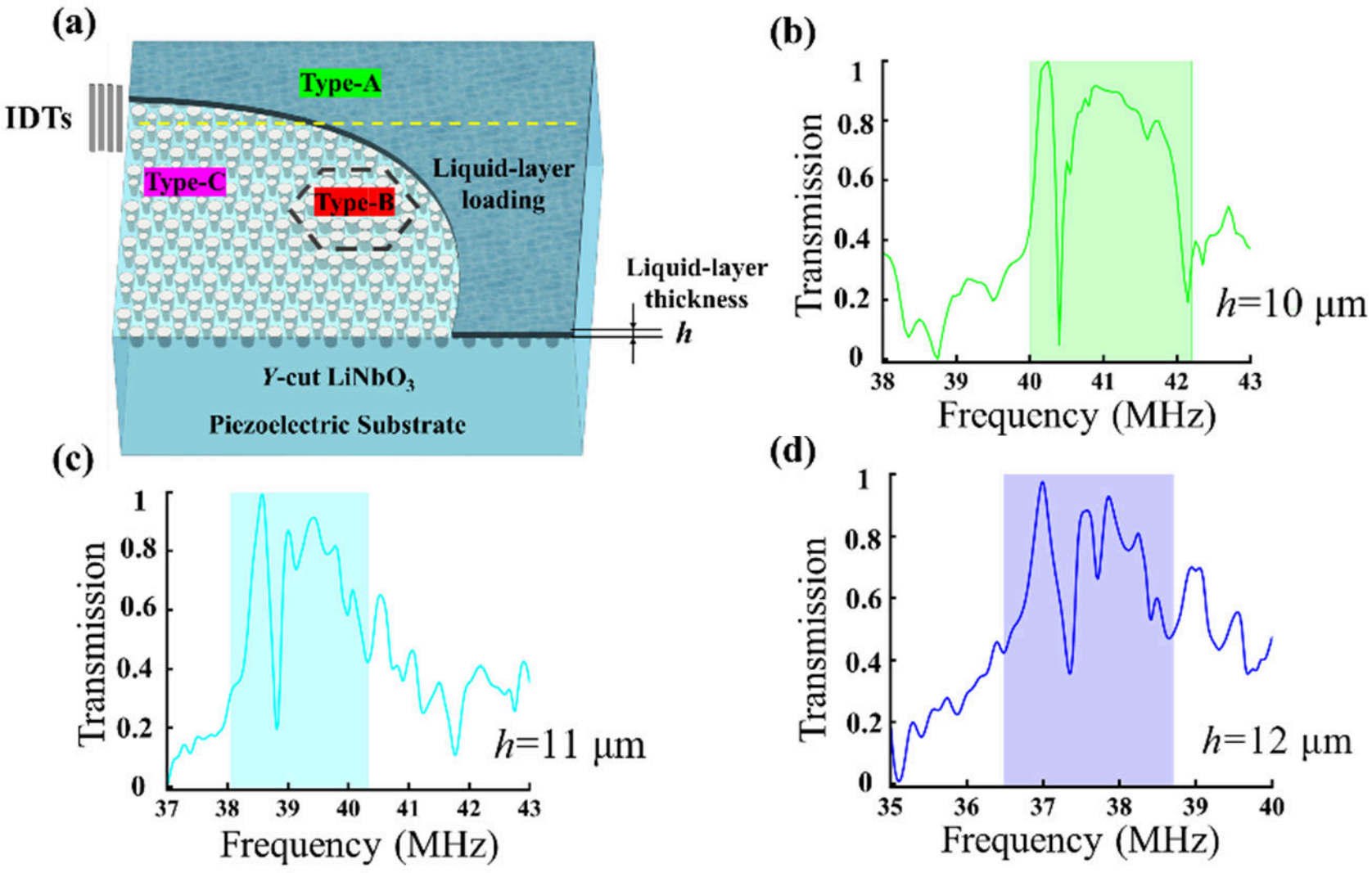

**Fig. 10.** The influences of the liquid-layer thickness on characteristics of the topological resonance.

(a) The coupling system including a straight waveguide and a resonant cavity under a liquid-layer loading. (b)-(d) Normalized transmission spectrums of the coupling systems with different liquid-layer thicknesses.

Finally, the influences of the coupling strength between the waveguide and resonant cavity on characteristics of the topological resonance are analyzed. Specifically, three different coupling distances are set: $d$ =1 layer, $d$ = 2 layers, and $d$=3 layers, where a layer represents the lattice constant of the system. The results are shown in Fig. 11. It is shown that as the coupling distance between the waveguide and the resonant cavity increases, notable decrease in Q-factor of the resonance peak emerges. The larger the coupling distance is, the weaker the coupling strength is, resulting in a lower Q-factor. The results indicate that the highest Q-factor can be achieved when the distance is $d$ = 1 layer.

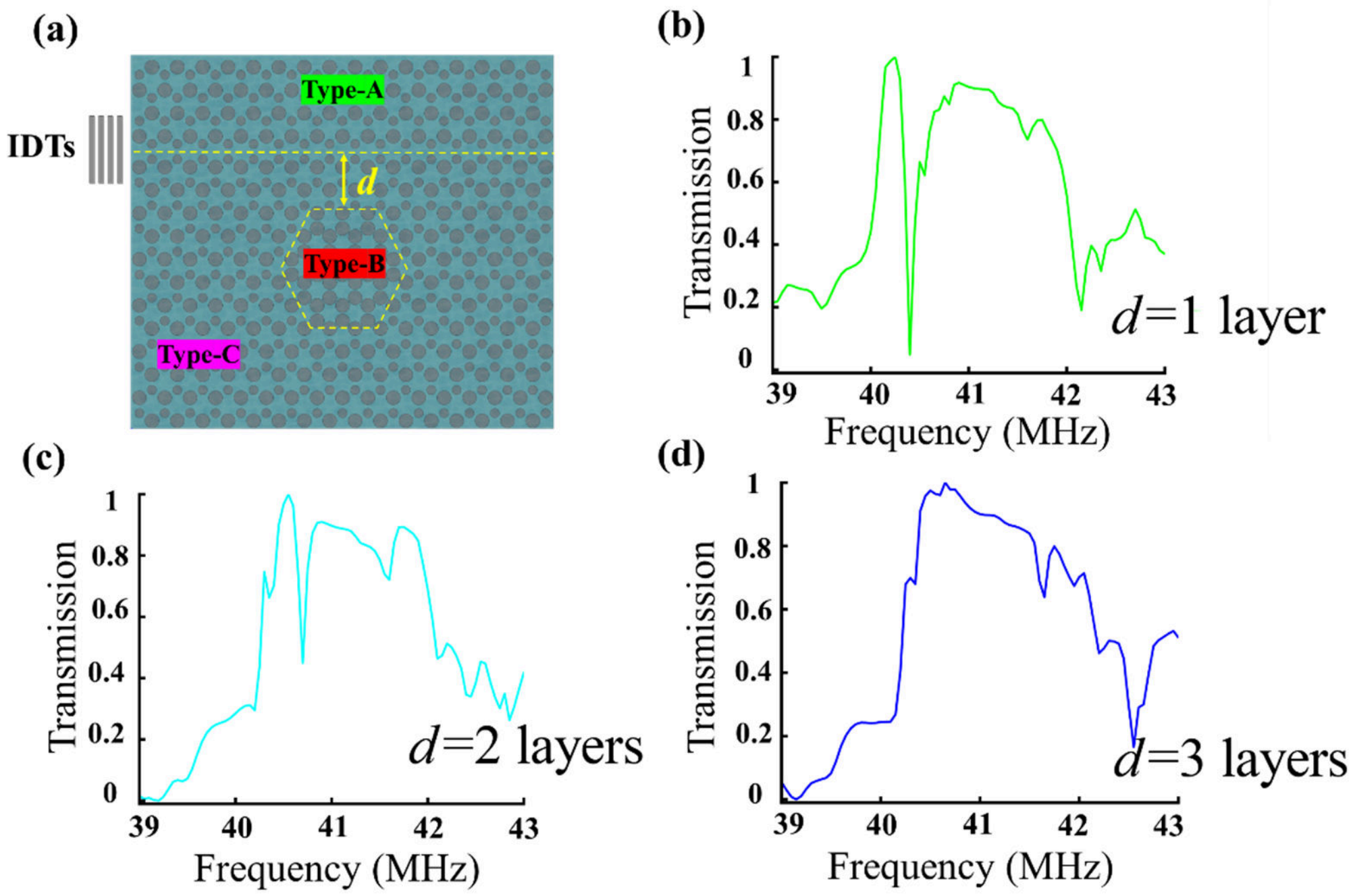

**Fig. 11.** The influences of the coupling distance on characteristics of the topological resonance. (a) a coupling system including a straight waveguide and a resonant cavity under a liquid-layer loading. (b-d) Normalized transmission spectrums of systems with different coupling distances ($d$=1 layer, $d$=2 layers, and $d$=3 layers).

## 4. Sensing characteristics

The topological resonance sensor is used to detect the NaCl concentration in aqueous solutions. Transmission spectrums of the devices with varying NaCl concentrations are shown in Fig. 12 (a). The relationship between the resonance frequency and the concentration of NaCl in aqueous solutions is shown in Fig. 12 (b), the concentration ranges from 0% to 5%.

Several performance parameters of the sensor are calculated, including sensitivity (S), quality factor (Q), and figure of merit (FOM). For a sharper resonance peak, the value of quality factor (Q) is higher, which can result in a higher frequency resolution; The sensitivity (S) represents the relationship between the shift of the resonance frequency and the change of the analyte concentration; The figure of merit (FOM) indicates the quality of sensor to measure any change in the resonant frequency. The detail definitions of these parameters are presented in **Appendix F**.

Table 2 presents the sensing performance of the topological resonance sensor under solution-layer loadings with different NaCl concentrations in aqueous solutions. The results show that within NaCl concentration range of 0~5%, the sensitivities are from 142.71 to 157.52 kHz/%. A linear fitting of the resonance frequency vs. NaCl concentration is also carried out, as shown in Fig. 12 (b). The relationship can be described by the following equation:

$$F = 0.14889c + 40.46495, \quad R^2 = 0.99966.$$

The above equation indicates that for 1% change in NaCl concentration, the resonance frequency increases by 148.89 kHz. The fitted $R^2$ value is 0.99966. Here R2 is the fit measure of the linear fitted regression model, and the fit close to 1 indicates an excellent fit. In the inset of Fig. 12 (b), in order to assess the resolving ability of the device, NaCl solutions with concentrations of 0.001 % and 0.002 % are selected, and it can be seen that the resonance frequency shift of the device is 160 Hz.

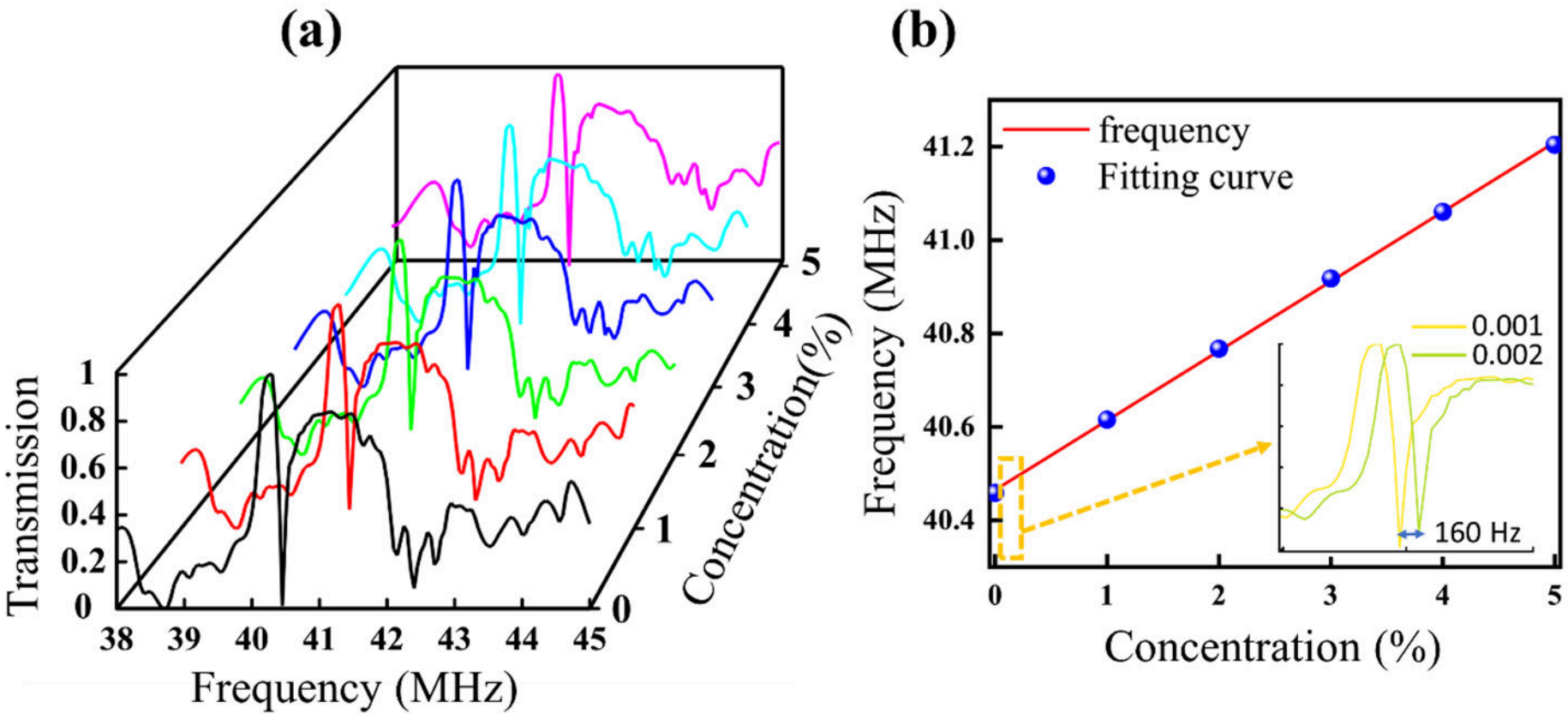

**Fig. 12.** Sensing characteristics of the SAW topological resonance device sensing NaCl concentration in aqueous solutions. (a) Transmission spectrums of the device with varying NaCl

concentrations, ranging from 0% to 5%. (b) Relationship between the resonance frequency and the NaCl concentration.

**Table 2.** Performance parameters of the topological resonance sensor for detecting NaCl concentrations in aqueous solutions

| Concentration (%) | Frequency (MHz) | Q-factor | Sensitivity (kHz/%) | FOM |
|---|---|---|---|---|
| 0 | 40.458 | 434.895 | - | - |
| 1 | 40.616 | 399.801 | 157.52 | 1.551 |
| 2 | 40.770 | 394.177 | 153.91 | 1.488 |
| 3 | 40.918 | 408.116 | 148.05 | 1.477 |
| 4 | 41.061 | 474.742 | 142.71 | 1.650 |
| 5 | 41.204 | 464.215 | 143.29 | 1.614 |

The topological resonance sensor with a straight waveguide is used to detect the NaI concentration in aqueous solutions. Transmission spectrums of the straight waveguide with varying NaI concentrations are shown in Fig. 13 (a), the relationship between the resonance frequency and the concentration of NaI in aqueous solutions is shown in Fig. 13(b). Table 3 presents the sensing performance of the topological resonance sensor for detecting NaI concentration. The results show that within NaI concentration range of 0~5%, the sensitivities are from 43.0 to 54.44 kHz/%. A linear fitting of the resonance frequency is also carried out, as shown in Fig. 13 (b). The relationship can be described by the following equation:

$$F = 0.04865c + 39.69749, \quad R^2 = 0.99828.$$

The equation indicates that for 1% change in NaI concentration, the resonance frequency increases by 48.65 kHz. The fitted $R^2$ value is 0.99828. In the inset of Fig. 13(b), in order to assess the resolving ability of the device, NaI solutions with concentrations of 0.001 % and 0.002 % are selected, and it is shown that the resonance frequency shift of the device is 60 Hz.

Table 4 presents a comparison between sensing performances of the proposed topological resonance sensor in this work and those of acoustic sensors from previous research reports. It is shown that the sensitivities in this work is higher obviously than those of previously reported sensors.

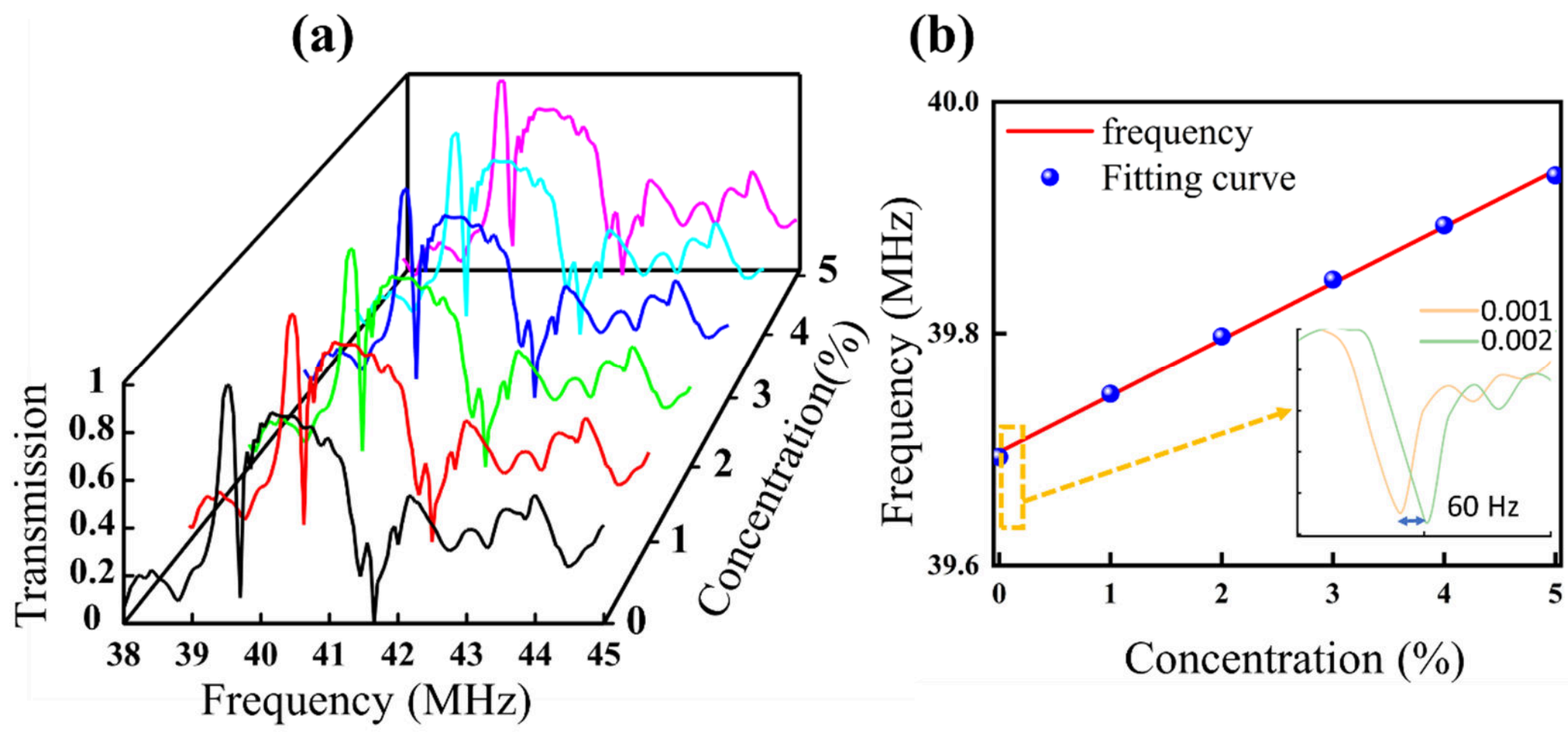

**Fig. 13.** Sensing characteristics of the SAW topological resonance device sensing NaI concentration in aqueous solutions. (a) Transmission spectrums of the device with varying NaI concentrations ranging from 0% to 5%. (b) Relationship between the resonance frequency and the NaI concentrations.

**Table 3.** Performance parameters of the topological resonance sensor for detecting NaI concentrations in aqueous solutions

| Concentration (%) | Frequency (MHz) | Q-factor | Sensitivity (kHz/%) | FOM |
|---|---|---|---|---|
| 0 | 39.693 | 475.539 | - | - |
| 1 | 39.748 | 405.340 | 54.44 | 0.555 |
| 2 | 39.797 | 463.299 | 49.75 | 0.579 |
| 3 | 39.847 | 428.000 | 49.11 | 0.527 |
| 4 | 39.893 | 377.672 | 46.95 | 0.445 |
| 5 | 39.937 | 332.140 | 43.00 | 0.358 |

**Table 4.** Comparison of the sensitivity of acoustic sensors

| Reference | Detection object | Sensitivity |
|---|---|---|
| Hamed Gharibi et al. [57] | NaI–water | 2177 Hz/% |
| Abdellatif Gueddida et al. [58] | NaI–water | 566 Hz/% |
| Frieder Lucklum et al.[59] | NaCl in water | 1441 Hz/% |
| LuLu Liang et al. [60] | NaCl in water | 42 kHz/% |
| Our work | NaCl in water<br>NaI in water | 148.89 kHz/%<br>48.65 kHz/% |

In addition, In **Appendix G**, the topological resonance sensor is used to detect hemoglobin and albumin concentrations in aqueous solutions.

## 5. Conclusion

Topological resonance behaviors of SAW under a surface liquid-layer loading are investigated, based on which, a novel liquid-phase sensor is proposed. The results show that the liquid-layer loading has an obvious influence on the degree of the spatial inversion symmetry, which can affect the coupling conditions between the topological waveguide and the resonant cavity, and topological indices (Berry curvature and Chern number). Based on that, a topological resonant peak with a high Q-factor is achieved based on couplings of a topological interface-mode waveguide and a resonant cavity under a surface liquid-layer loading.

The results show that the frequency of topological resonance peak is significantly sensitive to the liquid parameters. It is shown that the influences of the thickness of the liquid-layer loading on the resonance Q-factor is weak, thus this contributes greatly to the convenience of sensing experiments. Then, sensing performances of the topological resonance sensor are simulated. It is used to sense the concentration of hemoglobin, albumin, NaCl and NaI in aqueous solutions, and high sensitivities and Q-factors are obtained. The results presented in this paper can provide an important basis for the realization of highly sensitive and stable SAW micro-liquid-sample biomedical sensors in the future.

**Acknowledgments**

This work was supported by National Natural Science Foundation of China (No. 12172183), National Key Research and Development Program of China (No. 2023YFE0111000), the Natural Science Foundation of Zhejiang Province (No. LZ24A020001), International Science and Technology Cooperation Project lauched by Science and Technology Bureau of Ningbo City, Zhejiang Province, China (No. 2023H011), One health Interdisciplinary Research Project (No. HY202206), Ningbo University. This work was partially supported by Russian Ministry of Science and Higher Education (Contract #075-15-2023-580).

# Appendix A. Influences of surface liquid-layer loadings on dispersion band structure of phononic crystals

The simulated dispersion band structures for the phononic crystal (for the unit cell, $r_A = 3.9$ μm, $r_B = 7$ μm ) with and without surface liquid-layer loadings are shown in Fig. A.1. The blue dots represent the band structure for that with no surface liquid-layer loading, while the red dots represent that with a surface water-layer loading of 10μm thickness. The result show that after the surface of the structure is covered with a water layer, due to the fluid-solid coupling effect, many additional bands (red dots) emerge.

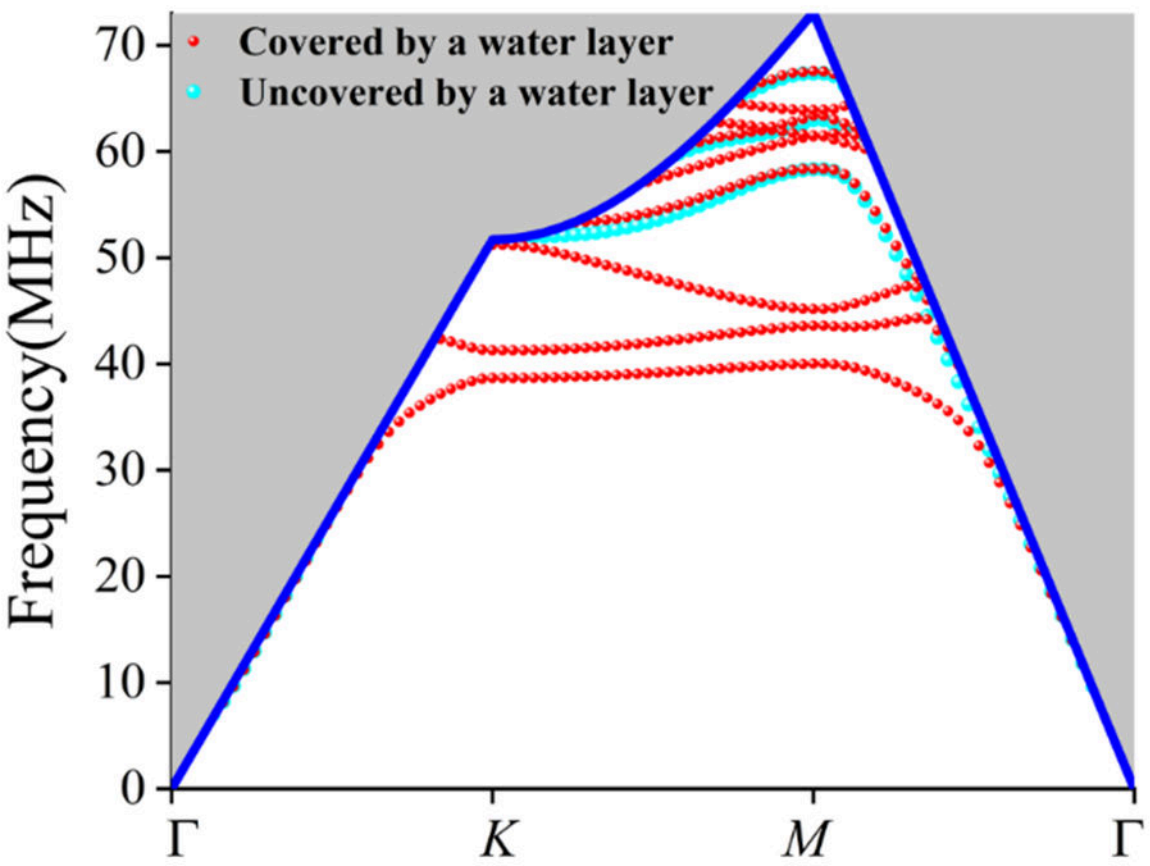

**Fig. A.1.** The dispersion band structures of the phononic crystal (for the unit cell, $r_A = 3.9$ μm, $r_B = 7$ μm ). The blue dots represent the band structure for that with no surface water-layer loading, while the red dots represent that with a surface water-layer loading of 10μm thickness.

# Appendix B. The complete dispersion band structures

The complete dispersion band structures of SAW phononic crystals with different unit cell types are shown in Fig. B.1. This work focuses on the first and second bands.

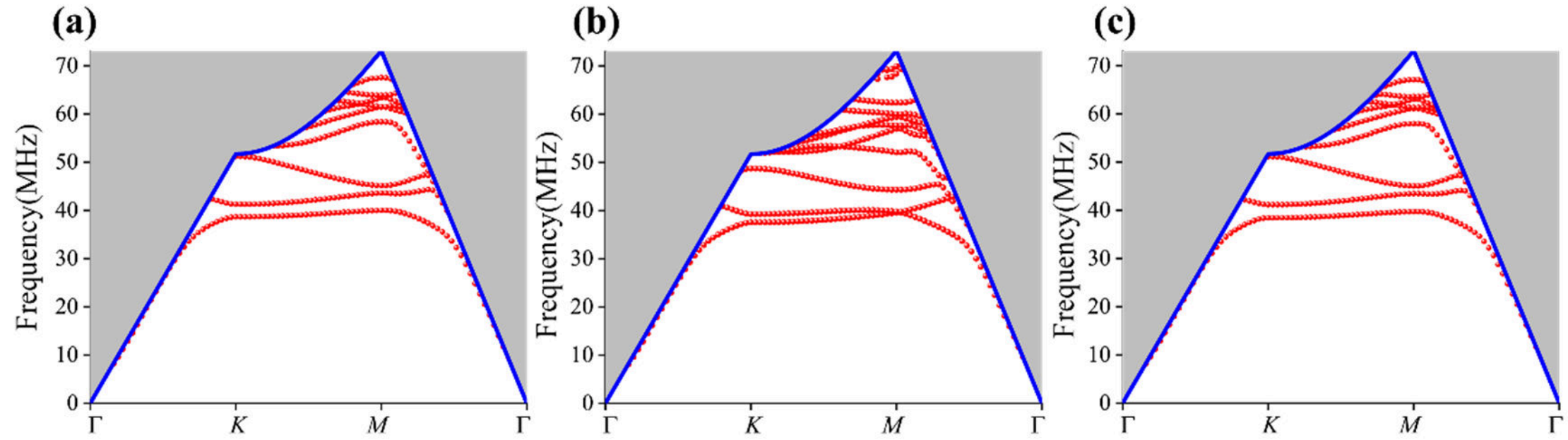

**Fig. B.1.** The complete dispersion band structures of SAW phononic crystals shown in Fig.4. (a) Type-A unit cell. (b) Type-B unit cell. (c) Type-C unit cell. The red dots represent the SAW modes, and the gray regions indicate the sound cone.

## Appendix C. Recognition of SAW modes

It is necessary to distinguish SAW modes from bulk acoustic wave (BAW) modes. For each eigenmode, the ratio $L$ of the elastic energy on the model surface to that inside the model can be calculated as:

$$L = (\iiint_{surface} \mathrm{EdV}) / (\iiint_{all} \mathrm{EdV}).$$

When $L > 0.8$, the majority of the elastic energy of the phononic crystal exists on the surface, indicating it belongs to SAW mode. When $L$ is lower, most of the energy exists within the interior of the model, indicating that it belongs to BAW mode. When the unit cell model has sufficient depth, this approach is effective [61,62]. Meanwhile, the sound lines are plotted to further distinguish SAW and BAW modes, as shown in Fig. C.1, where the red dots represent SAW modes and the black dots represent BAW modes.

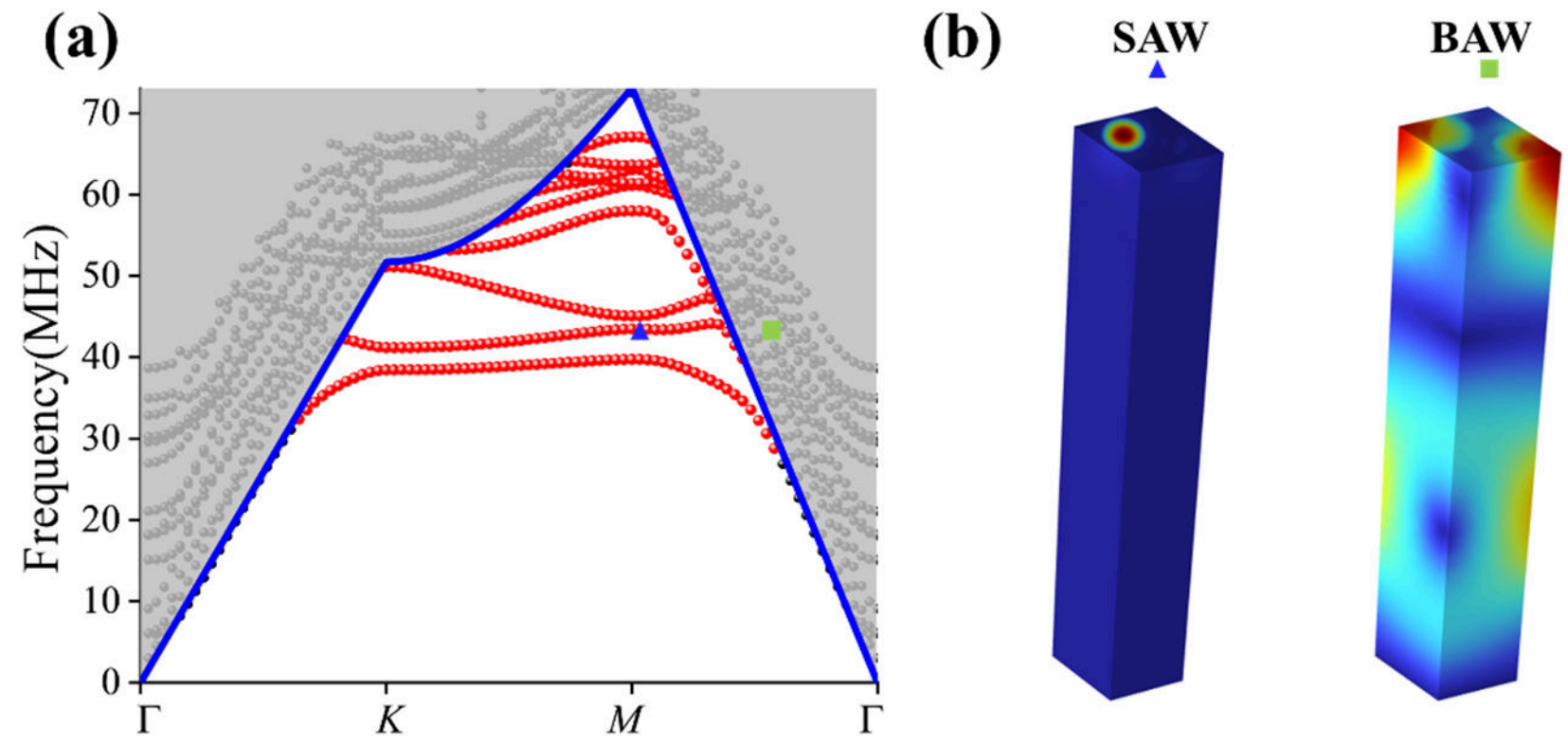

**Fig. C.1.** The dispersion band structure of SAW phononic crystals shown in Fig.4. (a) The band structure of the type-A unit cell, the red dots represent SAW modes (L>0.8), and the black dots represent BAW modes (L<0.3). (b) The elastic energy distribution of SAW and BAW modes, the blue triangle represents the typical SAW mode, and the green square represents the typical BAW mode.

To further confirm that the SAW mode exists, we also adopted the method in references [63], based on the criteria of energy distributions and the depth of energy penetration. Total displacement with depth is used to describe the energy distribution characteristics of SAW. The result is shown in Fig. C.**2**. It can be seen that for the depth of 1.4 λ， the total displacement is smaller than one ten-billion of its maximum value. It is shown that the displacement decays rapidly with the increasing depth, which verifies that the mode studied in this work is the SAW mode.

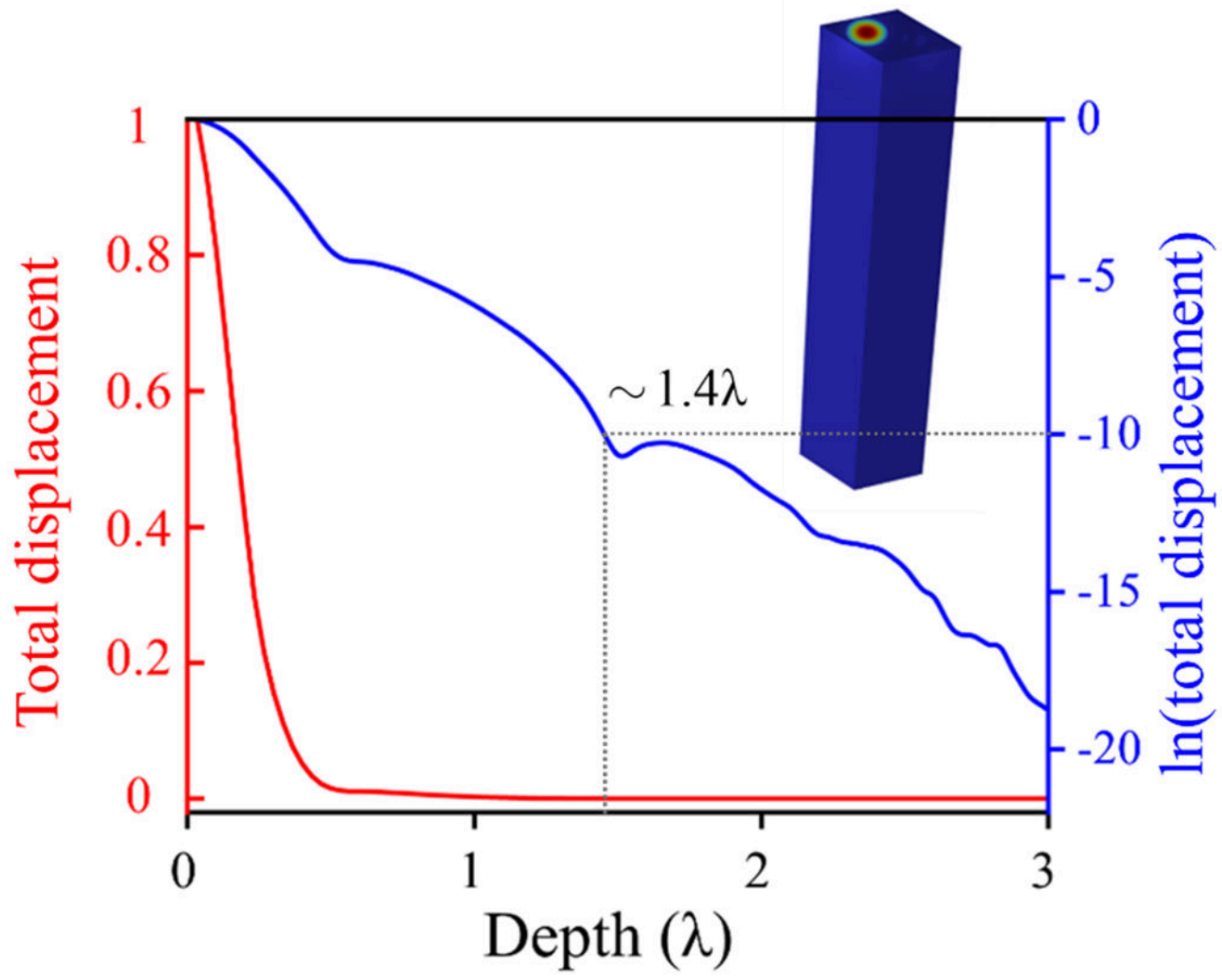

**Fig. C.2.** The variation of the total displacement amplitude with depth.

## Appendix D. Calculation of Berry curvature

For numerical discretization, $\delta k_x = \delta k_z = 4.18\ \mathrm{e}^3\mathrm{m}^{-1}$ (as shown in Fig. D.1), Berry curvature $\Omega(\mathrm{k})$ can be rewritten as

$$\Omega(\mathbf{k}) = \mathrm{Im}\,\mathrm{In}\,[\boldsymbol{U}^{(n)}_{\mathbf{k}_1\rightarrow\mathbf{k}_2}\boldsymbol{U}^{(n)}_{\mathbf{k}_2\rightarrow\mathbf{k}_3}\boldsymbol{U}^{(n)}_{\mathbf{k}_3\rightarrow\mathbf{k}_4}\boldsymbol{U}^{(n)}_{\mathbf{k}_4\rightarrow\mathbf{k}_1}],$$

where $\boldsymbol{U}^{(n)}_{\mathbf{k}_\alpha\rightarrow\mathbf{k}_\beta}$ denotes the displacement field in $y$-direction of the eigenmode, which is calculated by using COMSOL，namely, $\boldsymbol{U}^{(n)}_{\mathbf{k}_\alpha\rightarrow\mathbf{k}_\beta} = \dfrac{\langle \mathbf{u}(\mathbf{k}_\alpha) | \mathbf{u}(\mathbf{k}_\beta)\rangle}{|\langle \mathbf{u}(\mathbf{k}_\alpha) | \mathbf{u}(\mathbf{k}_\beta)\rangle|}$, $\alpha,\beta = 1,2,3,4$ and $\mathbf{k}_1, \mathbf{k}_2, \mathbf{k}_3, \mathbf{k}_4$ vertices represent the wavevectors at the black dots in Fig. D.1.

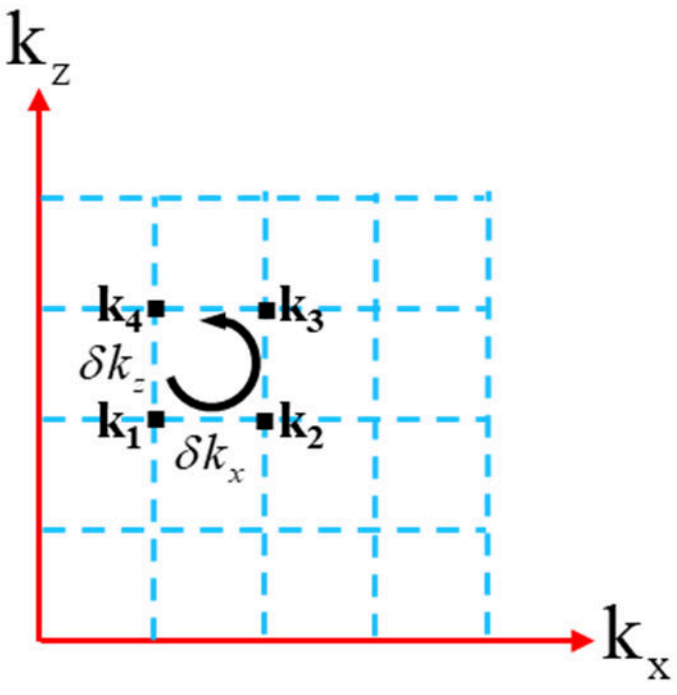

**Fig. D.1.** Calculation schematic of Berry curvature.

Here, we calculated the Berry curvature distribution at the $\mathrm{M}$ and $\mathrm{M}'$ points of the phononic crystal when the symmetry breaking degree gradually increases, as shown in Fig. D.2. The bandgap is gradually widened with the increasing symmetry-breaking degree, while the intensity and localization of the Berry curvature around the $\mathrm{M}$ and $\mathrm{M}'$ points gradually decrease, and the Chern number also decreases obviously. Here, normalized processing is carried out for the calculation of the Berry curvature.

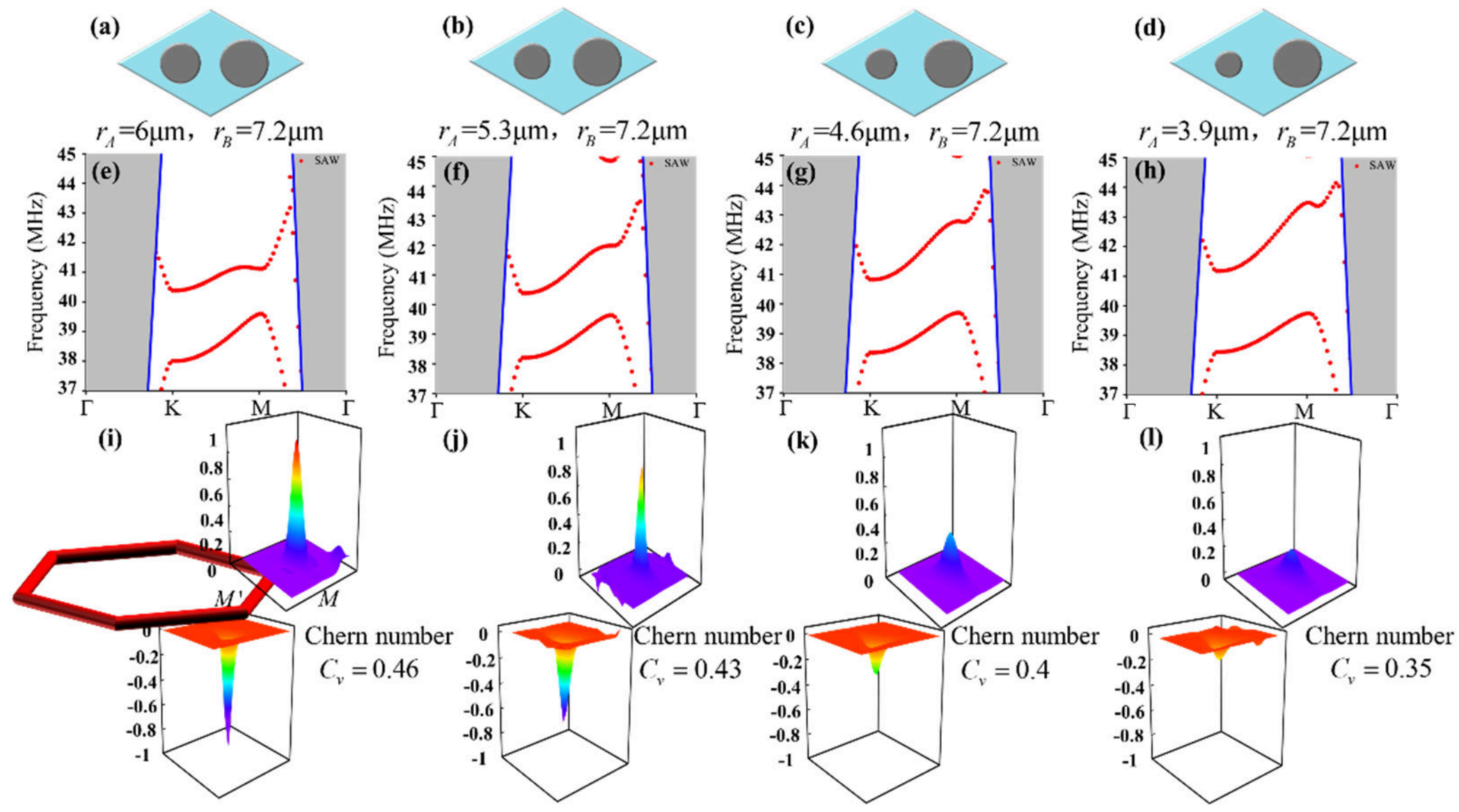

**Fig. D.2.** The band structures, Berry curvatures and Chern numbers corresponding to different unit cells. (a)-(d) The unit cells with the increasing symmetry-breaking degree (increasing radius difference of the two columns). (e)-(h) The bandgaps in the band structure of the unit cells shown in (a)-(d). The bandgap is gradually widened with the increasing symmetry-breaking degree. (i)-(l). The distribution of the Berry curvature and Chern number of the unit cells shown in (a)-(d). The intensity and localization of the Berry curvature around the M and M' points and chern number gradually decrease with the increasing symmetry-breaking degree.

# Appendix E. Topological transmissions of SAW under a liquid-layer loading

The supercell is shown in Fig. E.1(a), where the left is A-type phononic crystal, the right is C-type phononic crystal. The phononic crystals are covered by a water layer of 10 μm thickness. Firstly, numerical simulations are performed for the interface states of the supercell. Periodic boundary conditions are applied on both sides of the supercell, and the calculated dispersion relations are shown in Fig. E.1(c). It is shown that when the thickness of the water layer is 10 μm, valley Hall interface states (marked by red squares) is observed in the frequency range of 40.93-42.5 MHz. Besides, it can be seen that the energy is mostly concentrated near the A-C interface.

Fig. E.1(d) shows interface modes of supercells under water-layer loadings with different thicknesses. The results show that with the decreasing (increasing) thickness of the liquid-layer loading, the operational frequency of the interface state gradually shifts towards higher (lower) frequencies. This expands the range of operating frequencies of phononic-crystal sensors. Thus, the frequency of the interface state can be tuned conveniently by adjusting the thickness of the liquid layer. However, with an excessive increase (or decrease) of liquid-layer thickness, the interface state will mix with ordinary modes. This prevents energy from concentrating at the interface, thereby affecting the SAW transmission in the waveguide.

In order to further confirm the effect of different liquids on the interface mode, water (speed of sound is $c$=1500 m/s, density is $\rho$=1000 kg/m$^3$), olive oil (speed of sound is $c$=1431 m/s, density is $\rho$=910 kg/m$^3$), kerosene (speed of sound is $c$=1324 m/s, density is $\rho$=810 kg/m$^3$), petroleum (speed of sound is $c$=1290 m/s, density is $\rho$=876 kg/m$^3$), gasoline (speed of sound is $c$=1250 m/s, density is $\rho$=700 kg/m$^3$) and ethyl alcohol (speed of sound is $c$=1207 m/s, density is $\rho$=790 kg/m$^3$) are selected to cover the phononic crystals, respectively, the according interface modes are shown in Fig.E.1(e). The results show that as the material of the liquid layer changes, the frequency of the interfacial state changes obviously.

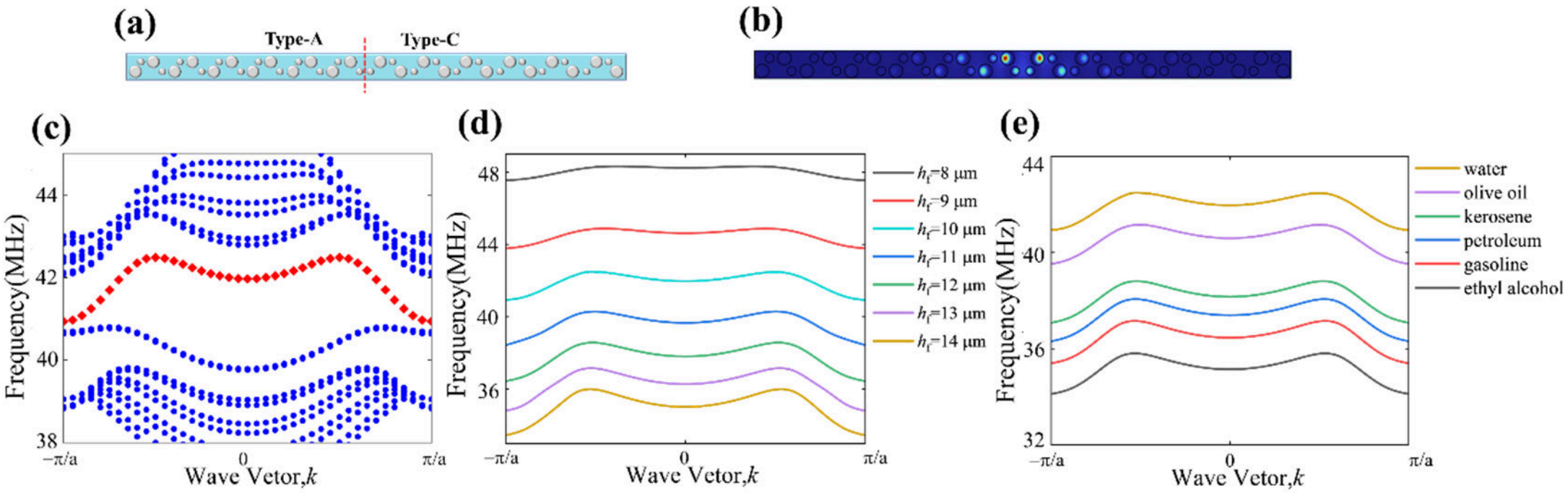

**Fig. E.1** (a) The schematic diagram of the supercell, which is composed of two types of unit cells, Type-A and Type-C, with an interface in the middle. (b) The displacement distribution of the interface states. (c) The elastic band structure of the supercell, red squares represent the interface states. (d) The evolution of the dispersion curve of the interface state with the change of the liquid-

layer thickness. As the thickness of the liquid-layer increases, the frequency of the interface state gradually decreases. (e) The evolution of the dispersion curve of the interface state with the change of the liquid-layer material. Here, water, olive oil, kerosene, petroleum, gasoline and ethyl alcohol are selected, respectively.

Numerical simulations are carried out to investigate the robustness of the interface mode to potential manufacturing defects. For a straight waveguide, as shown in Fig.E.2 (a), the transmission performance is good when no defect or disorder is introduced. Then, a defect (two pillars are deleted in straight path) and disorder（the radius of the two pillars on the straight path has been changed from 3.9μm to 5.3μm）are introduced at the transmission path of the interface states. The transmission results are shown in Fig. E.2 (b-c), respectively. It can be seen that for the defect and disorder, there is no significant scattering or energy attenuation, indicating that the SAW topological transmissions under surface liquid-layer loading are immune to the defects and disorders.

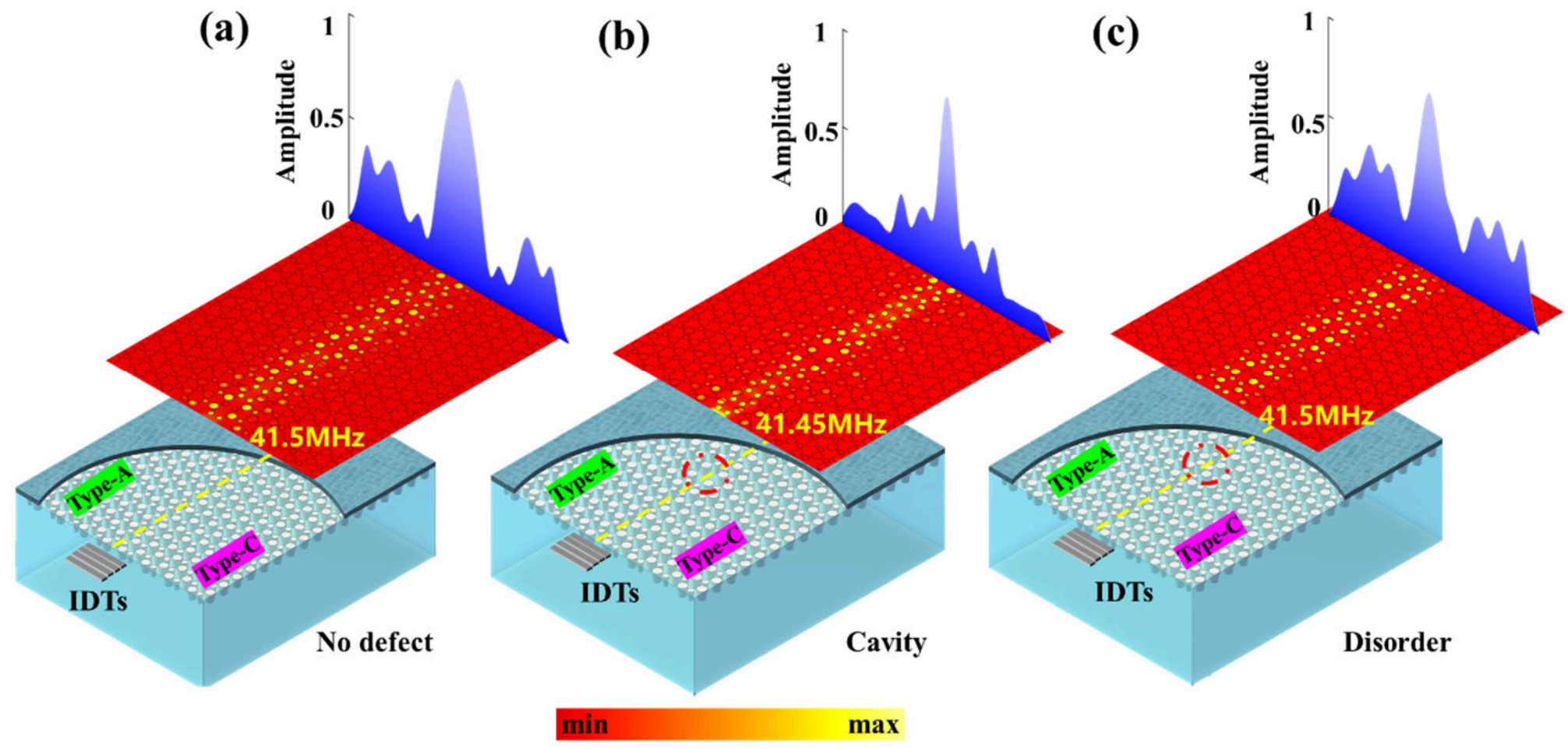

**Fig. E.2.** The displacement distributions of SAW along a straight topological interface waveguide under a liquid-layer loading. (a) With no defect. (b) With a defect (the two pillars are deleted). (c) With a disorder (The radius of the two pillars has been changed from 3.9μm to 5.3μm). The blue curve in the figure represents the amplitude distribution at the endpoint of SAW transmission.

# Appendix F. Sensing parameters

The quality factor (Q-factor) is defined as the ratio of the resonance frequency to the width at half maximum of the resonance transmission spectrum, namely

$$Q = \frac{f_r}{f_{HBW}}, \tag{F. 1}$$

where $f_r$ is the resonance frequency, and $f_{HBW}$ is the width at half maximum of the resonance transmission spectrum. The sharper the detected resonance peak is, the higher the Q-factor will be, which can increase the frequency resolution.

The sensitivity (S) is defined as the parameter representing the relationship between the shift of the resonance frequency and the concentration of the analyte, namely,

$$S = \frac{\Delta f}{\Delta C}, \quad \text{(F. 2)}$$

where $\Delta f$ represents the shift of the resonance frequency at different concentrations, and $\Delta C$ is the concentration change of the analyte.

The figure of merit (FOM) indicates the quality of sensor to measure any change in the resonant frequency. The calculation formula is

$$\text{FOM} = \frac{S}{f_{HBW}}. \quad \text{(F.3)}$$

# Appendix G. The detection of the hemoglobin and albumin concentrations in aqueous solutions

Firstly, the topological resonance sensor including a straight waveguide and resonant cavity is used to detect hemoglobin concentration in aqueous solution. Fig.G.1(a) shows transmission spectrums of the waveguide with varying hemoglobin concentrations, ranging from 0 to 0.14 mol/L. Fig.G.1(b) shows the relationship between the topological resonance frequency and the hemoglobin concentration. The results indicate that as the concentration of hemoglobin increases from 0 to 0.14 mol/L, the topological resonance frequency increases from 40.951 MHz to 41.487 MHz.

Table 5 shows values of performance parameters of the topological resonance sensor for detecting hemoglobin concentration in aqueous solutions. To further determine the relationship between sensitivity and concentration (molar ratio), a linear fit on the resonance frequency for each concentration (represented by blue dots in Fig.G.1(b)) is performed, as shown by the red line. This relationship can be described by the following equation:

$$F = 3.59552c + 40.29001, \quad R^2 = 0.99828.$$

From the equation, it is evident that for a change of 1 mol/L in hemoglobin concentration, the resonance frequency increases by 3.59552 MHz. The fitted value of $R^2$ value is 0.99828.

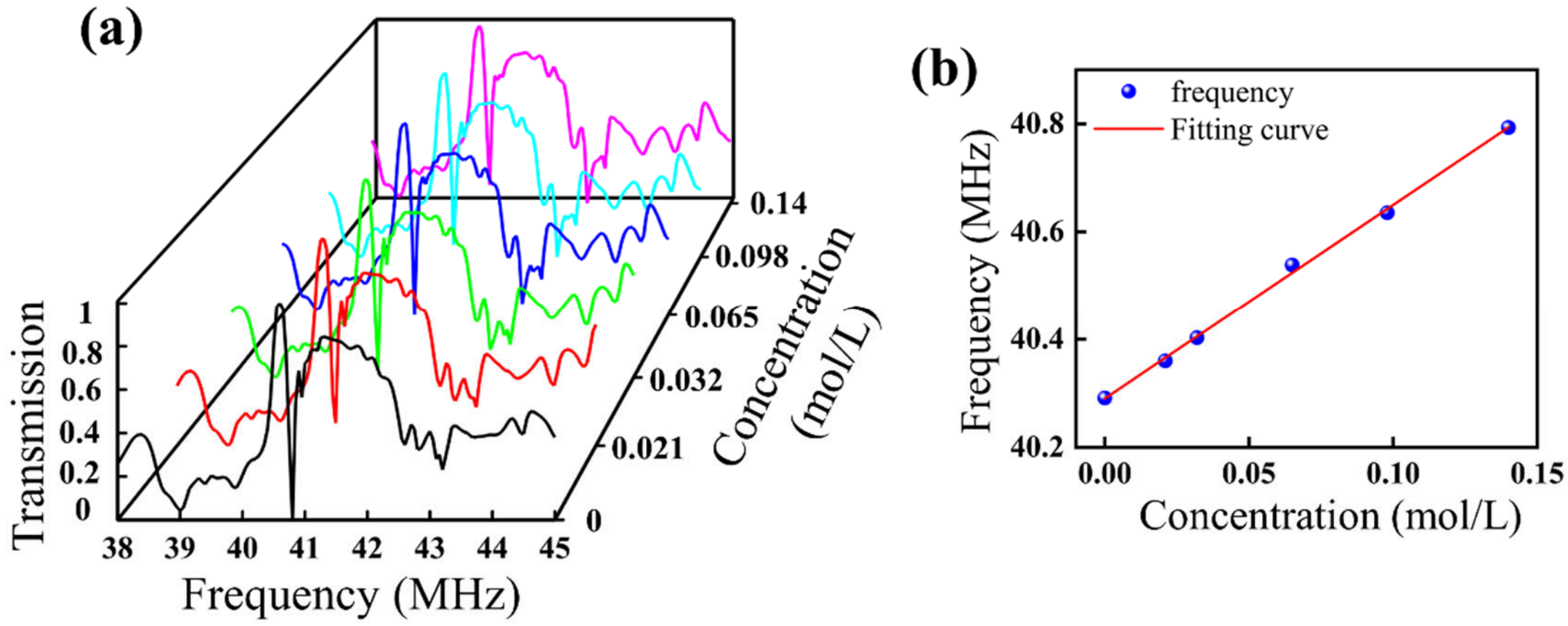

**Fig.G.1.** Sensing characteristics of the SAW topological resonance sensor for detecting hemoglobin concentrations in aqueous solutions. (a) Transmittance spectrums of the straight waveguide with various hemoglobin concentrations, ranging from 0 to 0.14 mol/L. (b) The relationship between the resonance frequency and the hemoglobin concentration.

**Table 5.** Performance parameters of the SAW topological resonance sensor for detecting hemoglobin concentrations in aqueous solutions

| Hemoglobin concentration (mol/L) | Resonance frequency (MHz) | Q-factor | Sensitivity (MHz/(mol/L)) | FOM |
|---|---|---|---|---|
| 0 | 40.291 | 413.156 | | |
| 0.021 | 40.360 | 432.213 | 3.286 | 35.186 |
| 0.032 | 40.403 | 472.771 | 3.909 | 45.742 |
| 0.065 | 40.538 | 453.801 | 4.091 | 45.795 |
| 0.098 | 40.635 | 333.593 | 2.939 | 24.131 |
| 0.14 | 40.793 | 390.364 | 3.762 | 36.000 |

Besides, the topological resonance sensor is used to detect albumin concentration in aqueous solutions. The concentration range of the solution is from 0 to 0.239 mol/L. The transmission spectrums are shown in Fig. G.2(a). The results show that the resonance frequency changes from 40.951 MHz to 42 MHz for the above concentration range. The performance parameters for different concentrations are listed in Table 6. A linear fit on the resonance frequency for each concentration (represented by blue dots in Fig.G.2(b)) is performed, as shown by the red line, and the relationship can be described by the following equation.

$$F = 3.26144c + 40.35755, \quad R^2 = 0.9965.$$

The equation indicates that for 1 mol/L change in albumin concentration, the resonance frequency increases by 3.26144 MHz. The fitted $R^2$ value is 0.9965.

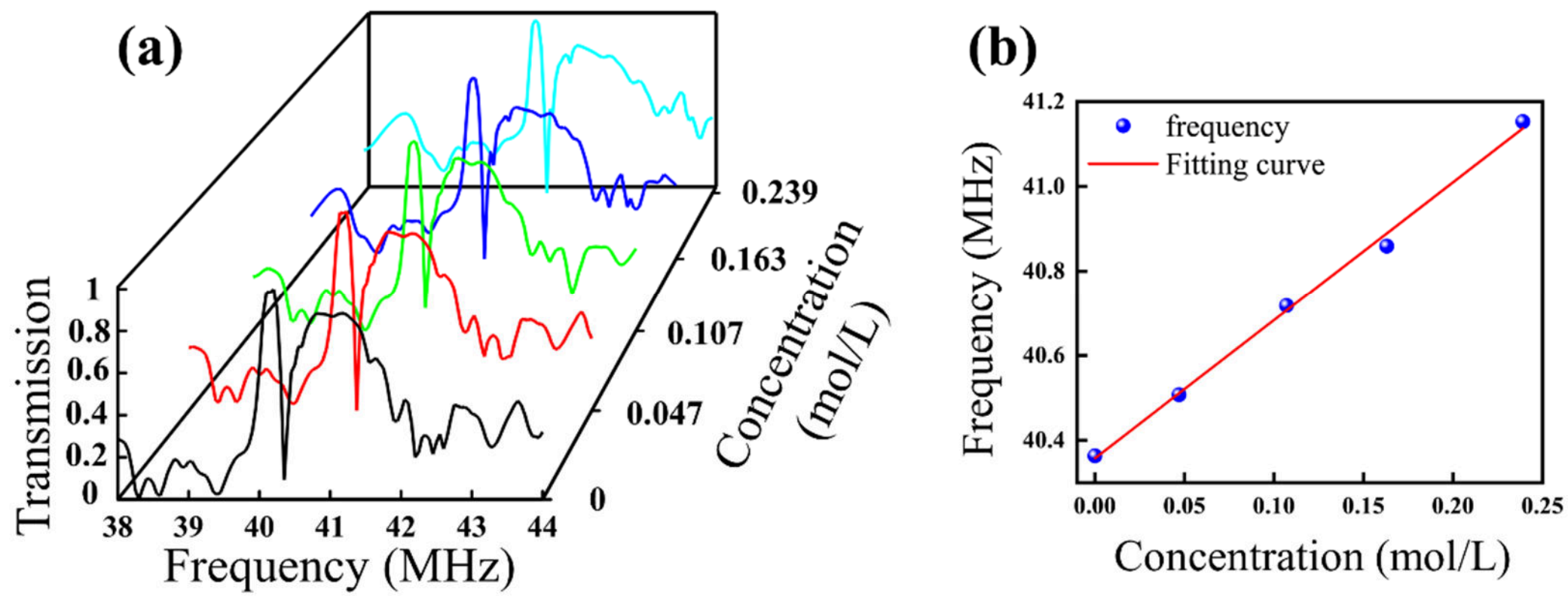

**Fig. G.2.** Sensing characteristics of the SAW topological resonance sensor for sensing albumin concentration in aqueous solutions. (a) Transmission spectrums of the straight waveguide with varying albumin concentrations, ranging from 0 to 0.239 mol/L. (b) Relationship between the resonance frequency and the albumin concentration.

**Table 6.** Performance parameters of the SAW topological resonance sensor for detecting albumin concentrations in aqueous solutions

| Concentration (mol/L) | Frequency (MHz) | Q-factor | Sensitivity (MHz/(mol/L)) | FOM |
|---|---|---|---|---|
| 0 | 40.364 | 423.409 | - | - |
| 0.047 | 40.507 | 479.150 | 3.059 | 36.178 |
| 0.107 | 40.718 | 440.432 | 3.510 | 37.963 |
| 0.163 | 40.859 | 495.745 | 2.524 | 30.627 |
| 0.239 | 41.153 | 477.192 | 3.865 | 44.817 |